\documentclass[12pt,english,reprint,amsmaths,amssymb,prl]{revtex4-1}
\usepackage[T1]{fontenc}
\usepackage[latin9]{inputenc}
\usepackage{color}
\usepackage{babel}
\usepackage{array}
\usepackage{refstyle}
\usepackage{booktabs}
\usepackage{multirow}
\usepackage{amsmath}
\usepackage{amssymb}
\usepackage{graphicx}
\usepackage[unicode=true,pdfusetitle,
 bookmarks=false,
 breaklinks=false,pdfborder={0 0 0},pdfborderstyle={},backref=false,colorlinks=true]
 {hyperref}

\makeatletter

\AtBeginDocument{\providecommand\secref[1]{\ref{sec:#1}}}
\AtBeginDocument{\providecommand\subsecref[1]{\ref{subsec:#1}}}
\AtBeginDocument{\providecommand\figref[1]{\ref{fig:#1}}}
\AtBeginDocument{\providecommand\tabref[1]{\ref{tab:#1}}}
\AtBeginDocument{\providecommand\eqref[1]{\ref{eq:#1}}}
\providecommand{\tabularnewline}{\\}
\newcommand{\lyxdot}{.}

\RS@ifundefined{subsecref}
  {\newref{subsec}{name = \RSsectxt}}
  {}
\RS@ifundefined{thmref}
  {\def\RSthmtxt{theorem~}\newref{thm}{name = \RSthmtxt}}
  {}
\RS@ifundefined{lemref}
  {\def\RSlemtxt{lemma~}\newref{lem}{name = \RSlemtxt}}
  {}

\renewcommand{\tabref}{\Tabref}
\renewcommand{\figref}{\Figref}
\renewcommand{\secref}{\Secref}
\usepackage[shortcuts]{extdash}

\makeatother

\begin{document}

\title{Theory of Ferroelectric ZrO$_{2}$ Monolayers on Si}

\author{Mehmet Dogan{*}$^{,1,2,3,4}$ and Sohrab Ismail-Beigi$^{1,2,5,6}$}

\affiliation{$^{1}$Center for Research on Interface Structures and Phenomena,
Yale University, New Haven, Connecticut 06520, USA $\linebreak$ $^{2}$Department
of Physics, Yale University, New Haven, Connecticut 06520, USA $\linebreak$
$^{3}$Department of Physics, University of California, Berkeley,
California 94720, USA $\linebreak$ $^{4}$Materials Science Division,
Lawrance Berkeley National Laboratory, Berkeley, California 94720,
USA $\linebreak$ $^{5}$Department of Applied Physics, Yale University,
New Haven, Connecticut 06520, USA $\linebreak$ $^{6}$Department
of Mechanical Engineering and Materials Science, Yale University,
New Haven, Connecticut 06520, USA $\linebreak$ {*}Corresponding author: mhmtdogan@gmail.com}

\date{January 31, 2019}
\begin{abstract}
We use density functional theory and Monte Carlo lattice simulations
to investigate the structure of ZrO$_{2}$ monolayers on Si(001).
Recently, we have reported on the experimental growth of amorphous
ZrO$_{2}$ monolayers on silicon and their ferroelectric properties,
marking the achievement of the thinnest possible ferroelectric oxide
{[}M. Dogan et al. \emph{Nano Lett., }\textbf{18 (1)} (2018) \citep{dogan2018singleatomic}{]}.
Here, we first describe the rich landscape of atomic configurations
of monocrystalline ZrO$_{2}$ monolayers on Si and determine the local
energy minima. Because of the multitude of low-energy configurations
we find, we consider the coexistence of finite-sized regions of different
configurations. We create a simple nearest-neighbor lattice model
with parameters extracted from DFT calculations, and solve it numerically
using a cluster Monte Carlo algorithm. Our results suggest that up
to room temperature, the ZrO$_{2}$ monolayer consists of small domains
of two low-energy configurations with opposite ferroelectric polarization.
This explains the observed ferroelectric behavior in the experimental
films as a collection of crystalline regions, which are a few nanometers
in size, being switched with the application of an external electric
field.
\end{abstract}
\maketitle

\section{Introduction\label{sec:Introduction}}

Thin films of metal oxides have been a focus area of continuous research
due to the rich physics that can be observed in these systems, such
as ferroelectricity, ferromagnetism and superconductivity, and their
resulting technological applications \citep{hwang2012emergent,mannhart2010oxideinterfacestextemdashan}.
An important challenge involving thin metal oxide films has been their
growth on semiconductors in such a way that their electrical polarization
couples to the electronic states inside the semiconductor \citep{reiner2009atomically,reiner2010crystalline,dogan2017abinitio}.
If successfully done, this enables the development of non-volatile
devices such as ferroelectric field-effect transistors (FEFET). In
a FEFET, the polarization of the oxide encodes the state of the device,
and requires the application of a gate voltage only for switching
the state, greatly reducing the power consumption and boosting the
speed of the device \citep{mckee2001physical,garrity2012growthand}.
Meeting this challenge requires a thin film ferroelectric oxide, as
well as an atomically abrupt interface between the oxide and the semiconductor,
so that the polarization of the oxide and the electronic states in
the semiconductor are coupled. The first of these requirements, i.e.,
a thin film ferroelectric, is difficult to obtain because materials
that are ferroelectric in the bulk lose their macroscopic polarization
below a critical thickness, due to the depolarizing field created
by surface bound charges \citep{batra1973phasetransition,dubourdieu2013switching}.
An alternative approach is to search for materials such that, regardless
of their bulk properties, they are stable in multiple polarization
configurations as thin films \citep{dogan2017abinitio}. The second
requirement, i.e., an abrupt oxide-semiconductor interface, has been
challenging due to the formation of amorphous oxides such as SiO$_{2}$
at the interface with a semiconductor such as Si \citep{robertson2006highdielectric,garrity2012growthand,mcdaniel2014achemical}.
This challenge has been overcome by using layer-by-layer growth methods
such as molecular beam epitaxy (MBE) and employing highly controlled
growth conditions \citep{mckee1998crystalline,mckee2001physical,kumah2016engineered}.

We recently reported on the experimental observation of polarization
switching in atomically thin ZrO$_{2}$ grown on Si \citep{dogan2018singleatomic}.
In the experimental setup, ZrO$_{2}$ was grown using atomic layer
deposition (ALD), yielding an amorphous oxide and an abrupt oxide-silicon
interface with no significant formation of SiO$_{2}$. This interface
was then incorporated into a gate stack device with amorphous Al$_{2}$O$_{3}$
separating it from the top electrode. Ferroelectric behavior was observed
by $C-V$ measurements with this gate stack. In this work, we present
an in-depth computational investigation of this monolayer system.

In \secref{Methods5}, we describe our computational methods. In \subsecref{Free-standing-ZrO2},
we investigate the structure of free-standing ZrO$_{2}$ monolayers
assuming they are strained to the two-dimensional lattice of the Si(001)
surface. In \subsecref{ZrO2-on-Si}, we report on the low-energy configurations
of these monolayers when placed on the Si(001) surface. We find that
these films have multiple (meta)stable structures with no significant
chemical differences between them. This suggests that epitaxial monocrystalline
growth may be challenging. In \subsecref{Domain}, we examine the
domain energetics in this system: we build a lattice model with nearest-neighbor
interactions, and solve this model using a Monte Carlo cluster method.
The results of the lattice model provide a microscopic understanding
of the experimentally observed polarization switching.

\section{Computational methods\label{sec:Methods5}}

\begin{figure}
\begin{centering}
\includegraphics[width=0.8\columnwidth]{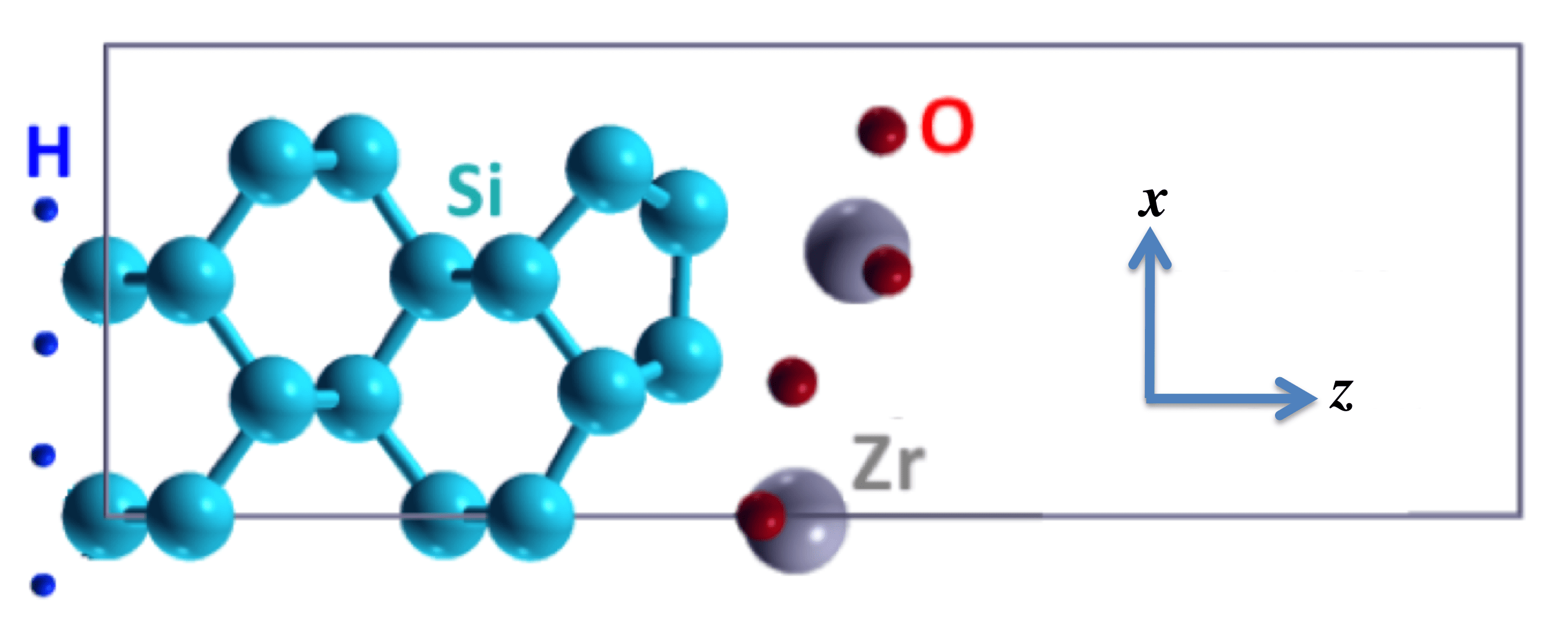}
\par\end{centering}
\caption[A typical simulation supercell with $2\times1$ in-plane periodicity.]{\label{fig:simcell}A typical simulation supercell with $2\times1$
in-plane periodicity. The bottom 4 layers of Si are fixed to bulk
coordinates and passivated by hydrogen as shown, to simulate bulk
silicon. There is $\sim12\text{\AA}$ of vacuum along the $z$-direction
to separate periodic copies.}
\end{figure}

We theoretically model the materials systems using density functional
theory (DFT) with the Perdew\--Burke\--Ernzerhof generalized gradient
approximation (PBE GGA) \citep{perdew1996generalized} and ultrasoft
pseudopotentials \citep{vanderbilt1990softselfconsistent}. We use
the QUANTUM ESPRESSO software package \citep{giannozzi2009quantum}.
A $35$ Ry plane wave energy cutoff is used to describe the pseudo
Kohn\--Sham wavefunctions. We sample the Brillouin zone with an $8\times8\times1$
Monkhorst\--Pack $k$-point mesh (per $1\times1$ in-plane primitive
cell) and a $0.02$ Ry Marzari\--Vanderbilt smearing \citep{marzari1999thermal}.
A typical simulation cell consists of $8$ atomic layers of Si whose
bottom layer is passivated with H and a monolayer of ZrO$_{2}$ placed
on top (see \figref{simcell}). Periodic copies of the slab are separated
by $\sim12\text{\AA}$ of vacuum in the $z$-direction. The in-plane
lattice constant is fixed to the computed bulk Si lattice constant
of $3.87\text{\AA}$. In general, the slab has an overall electrical
dipole moment along the $z$ direction that might artificially interact
with its periodic images across the vacuum. In order to prevent this
unphysical effect, we introduce a fictitious dipole in the vacuum
region of the cell which cancels out the electric field in vacuum
and removes such interactions \citep{bengtsson1999dipolecorrection}.
All atomic coordinates are relaxed until the forces on all the atoms
are less than $10^{-3}{\rm Ryd}/a_{0}$ in all axial directions, where
$a_{0}$ is the Bohr radius (except the bottom $4$ layers of Si which
are fixed to their bulk positions to simulate a thick Si substrate).
We use the nudged elastic bands (NEB) method with climbing images
\citep{henkelman2000aclimbing} to compute the transition energy barrier
between different metastable configurations.

\section{Results \label{sec:Results5}}

\subsection{Free standing ZrO$_{2}$ monolayers\label{subsec:Free-standing-ZrO2}}

\subsubsection{Background: bulk zirconia}

\begin{figure}
\begin{centering}
\includegraphics[width=0.8\columnwidth]{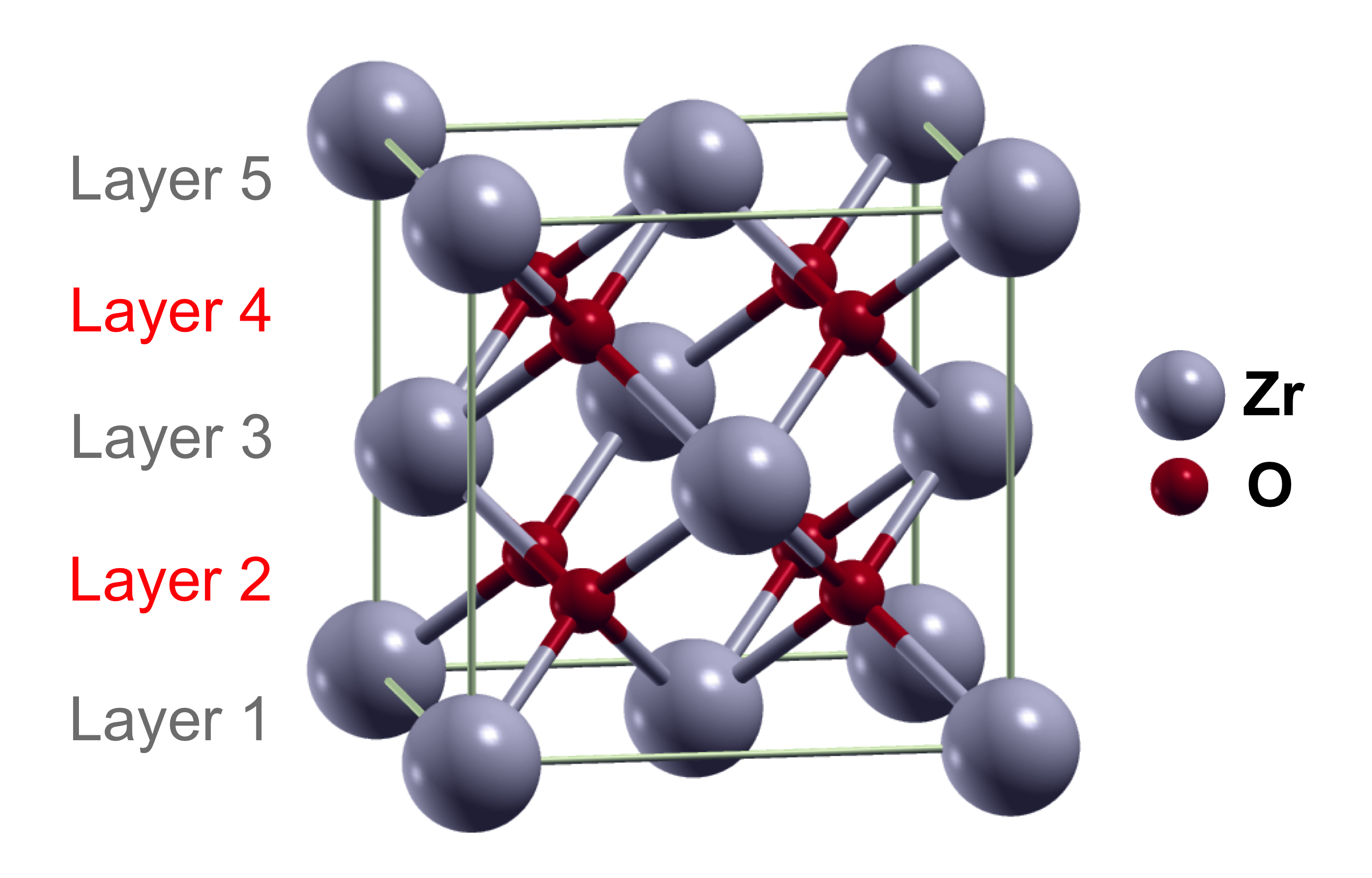}
\par\end{centering}
\caption[The high symmetry cubic phase ($Fm\overline{3}m$) of bulk ZrO$_{2}$. ]{\label{fig:cubicZrO2}The high symmetry cubic phase ($Fm\overline{3}m$)
of bulk ZrO$_{2}$. Atomic layers are labelled 1 through 5, where
the odd (even) layers correspond to cation (anion) planes.}
\end{figure}

Bulk ZrO$_{2}$ is observed in three structural phases. The high symmetry
cubic phase (space group: $Fm\overline{3}m$) is shown in \figref{cubicZrO2}.
The lower symmetry tetragonal ($P4_{2}/nmc$) and monoclinic ($P2_{1}/c$)
phases are obtained by continuously breaking the symmetries of the
cubic phase. All three configurations are centrosymmetric and hence
not ferroelectric. However, this binary oxide has a \emph{layered
structure} (along low-index directions) in which the cations and anions
lie in different planes, which, in thin film stoichiometric form,
would cause ultrathin ZrO$_{2}$ films to be polar. For instance,
in \figref{cubicZrO2} a horizontal monolayer of ZrO$_{2}$ could
be formed by the zirconium atoms in Layer 3, with (a) the oxygen atoms
in Layer 2, or with (b) the oxygen atoms in Layer 4, or with (c) half
of the oxygen atoms in each of Layer 2 and Layer 4. Before relaxing
the atoms in these hypothetical monolayers, in case (a) the resulting
polarization would be upward, in case (b) it would be downward, and
in case (c) it would be zero. This intrinsic layered structure, which
is also preserved in the tetragonal and the monoclinic phases of zirconia,
is a fundamental reason why ZrO$_{2}$ is an excellent candidate to
have a switchable polarization when grown on silicon.

\subsubsection{Structure of free standing monolayers}

In order to check if this richness of structure due to the layered
nature of the bulk material is retained in the ultrathin film, we
have simulated free standing ZrO$_{2}$ monolayers. A monolayer formed
by a (001) plane of cubic ZrO$_{2}$ would have a square lattice with
size $3.61\ \text{\AA}$ (based on our DFT computations). To match
the lattice of the Si substrate, we simulate the monolayers at the
lattice constant of the Si(001) surface, which we find to be $3.87\ \text{\AA}$.
We have searched for minimum energy configurations for $1\times1$,
$2\times1$, $2\times2$ and $c(4\times2)$ sized unit cells of monolayer
ZrO$_{2}$ which are the periodicities of the low energy reconstructions
of the bare Si(001) surface, as we shall discuss in \subsecref{ZrO2-on-Si}.

\begin{figure}
\begin{centering}
\includegraphics[width=1\columnwidth]{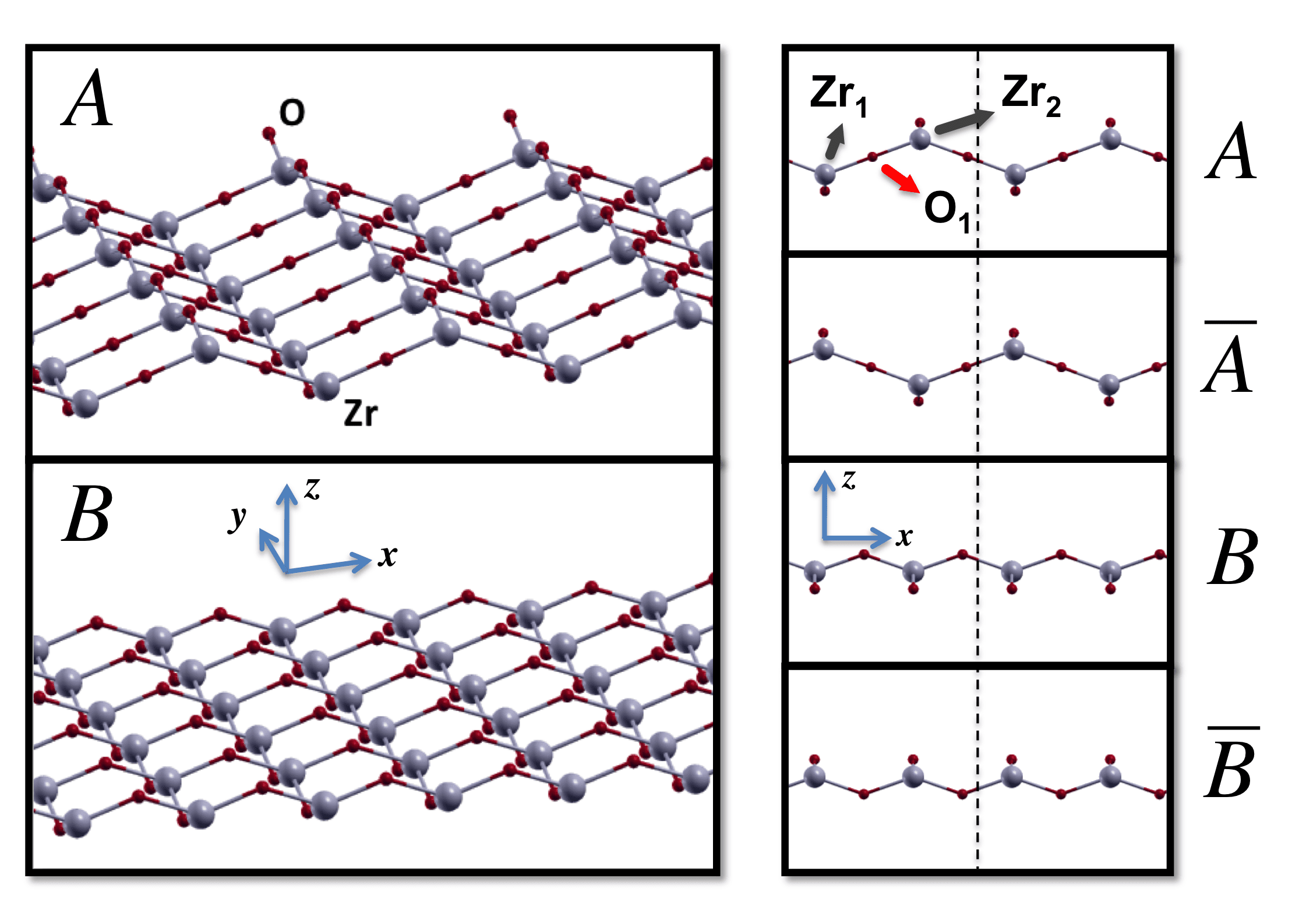}
\par\end{centering}
\caption[The lowest energy configurations of the free standing ZrO$_{2}$ monolayer.]{\label{fig:ZrO2_AB}The lowest energy configurations of the free
standing ZrO$_{2}$ monolayer. Structure $B$ has an energy of $0.07$
eV per ZrO$_{2}$ above that of structure $A$. On the right, all
four geometrically distinct metastable configurations are shown. $\overline{A}$
and $\overline{B}$ are obtained from $A$ and $B$, respectively,
by reflection in the $z=0$ plane. For each structure, two copies
of the $2\times1$ unit cells are displayed and a vertical dashed
line separates the copies.}
\end{figure}

We find that the lowest and the second lowest energy configurations
of the ZrO$_{2}$ monolayer are $2\times1$ and $1\times1$, respectively,
as shown in \figref{ZrO2_AB}. The chief difference between the two
configurations is that the lowest energy structure, labeled $A$,
has a vertical (along $z$) buckling of zirconiums in the $2\times$
in-plane direction, while for the second lowest energy structure,
labeled $B$, all the Zr are coplanar. We find that $E\left(B\right)-E\left(A\right)=0.07$
eV per ZrO$_{2}$. Both of these configurations are inversion symmetric
and hence non-polar. However, because neither $A$ or $B$ is symmetric
with respect to the mirror plane reflection $z\rightarrow-z$, there
are two more geometrically distinct minima, named $\overline{A}$
and $\overline{B}$, which are shown in \figref{ZrO2_AB}. $\overline{A}$
and $\overline{B}$ are obtained from $A$ and $B$, respectively,
by the mirror reflection. Notice that $\overline{A}$ can be obtained
from $A$ also by translating in the $2\times$ direction by half
a $2\times1$ cell. However, since the underlying substrate will have
at least $2\times1$ periodicity, this translation would not leave
the entire system (ZrO$_{2}$ with substrate) invariant.

\subsubsection{Energy landscape of free standing monolayers}

\begin{figure}
\begin{centering}
\includegraphics[width=1\columnwidth]{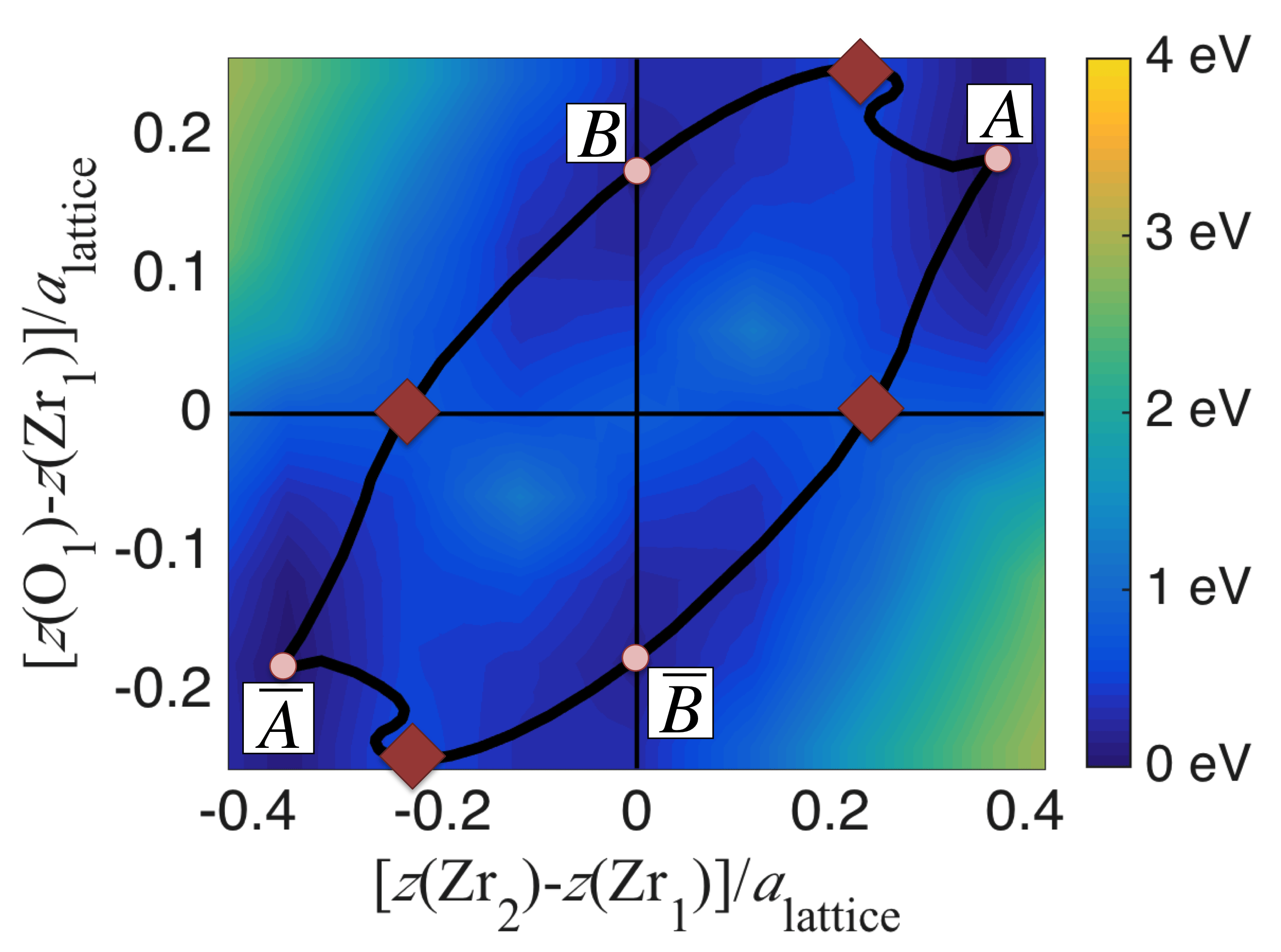}
\par\end{centering}
\caption[The energy landscape of the free standing ZrO$_{2}$ monolayer, as
parametrized by a pair of coordinates.]{\label{fig:ZrO2_barriers}The energy landscape of the free standing
ZrO$_{2}$ monolayer, as parametrized by a pair of coordinates $z_{1}\equiv z\left(\text{Zr}_{2}\right)-z\left(\text{Zr}_{1}\right)$
and $z_{2}\equiv z\left(\text{O}_{1}\right)-z\left(\text{Zr}_{1}\right)$
(See \figref{ZrO2_AB} for labelings of the atoms). $a_{\text{lattice}}$
is the computed lattice constant of silicon and is equal to $3.87\ \text{\AA}$.
All four local energy minima as well as the minimum energy transition
paths between them are shown. The saddle points on the landscape (i.e.,
the transition states) are shown as diamonds. The zero of energy is
taken to be the energy of structure $A$. All transition states lie
at the same energy because they are related by reflection/translation
operations. The energy landscape is computed by DFT on a $9\times9$
grid and then interpolated by MatLab to produce the smooth colored
plot.}
\end{figure}

In order to analyze these configurations further, we parametrize the
energy landscape of free standing ZrO$_{2}$ monolayers by using two
coordinates: $z_{1}\equiv z\left(\text{Zr}_{2}\right)-z\left(\text{Zr}_{1}\right)$
and $z_{2}\equiv z\left(\text{O}_{1}\right)-z\left(\text{Zr}_{1}\right)$,
where the atoms $\text{Zr}_{1}$, $\text{Zr}_{2}$ and $O_{1}$ are
labelled for structure $A$ in \figref{ZrO2_AB} (for structures $\overline{A}$,
$B$ and $\overline{B}$, Zr$_{1}$ is directly below Zr$_{1}$ of
structure $A$ in the figure, and similarly for Zr$_{2}$ and O$_{1}$).
Note that the structures $B$ and $\overline{B}$ are treated in $2\times1$
unit cells for this analysis. To explore the energy landscape, we
have made a $9\times9$ grid of $\left(z_{1},z_{2}\right)$ values
and computed corresponding energies for structures whose $z_{1}$
and $z_{2}$ are fixed but all other coordinates are relaxed. In \figref{ZrO2_barriers},
we plot the energy landscape using darker (lighter) colors to represent
lower (higher) energies. The coloring is implemented by MatLab's linear
interpolation scheme based on the DFT energies on an equally spaced
$9\times9$ grid. We also label the four (meta)stable configurations
on the landscape. The energies are reported for $2\times1$ cells
where $E\left(A\right)=E\left(\overline{A}\right)=0$ is set as the
zero of energy.

In \figref{ZrO2_barriers} we also present the minimum energy transition
paths between these energy minima, as thick solid curves. We have
found these transitions using the NEB method with climbing images
\citep{henkelman2000aclimbing}. There are 6 pairs of metastable configurations
and hence 6 transition paths: $A\leftrightarrow\overline{A}$, $A\leftrightarrow B$,
$A\leftrightarrow\overline{B}$, $\overline{A}\leftrightarrow B$,
$\overline{A}\leftrightarrow\overline{B}$ and $B\leftrightarrow\overline{B}$.
However, as seen from the figure, the transition paths of $A\leftrightarrow\overline{A}$
and $B\leftrightarrow\overline{B}$ go through other energy minima
and hence can be expressed in terms of the remaining 4 transitions.
We have found that all of the four transitions go through a transition
state with energy $1.04$ eV per $2\times1$ cell. These four saddle
points, shown as diamond marks in \figref{ZrO2_barriers}, are related
by reflection and/or translation operations, and hence are physically
equivalent.

To sum up, we have found that as a free standing monolayer in vacuum,
ZrO$_{2}$ is not polar but has two physically distinct stable configurations.
In the presence of a surface that breaks the $z\rightarrow-z$ symmetry,
$A$ and $\overline{A}$ (as well as $B$ and $\overline{B}$) have
the potential to relax to new configurations that are differently
polarized.

\subsection{ZrO$_{2}$ monolayers on Si(001)\label{subsec:ZrO2-on-Si}}

\subsubsection{Bare Si(001) surface}

To study the behavior of zirconia on Si(001), we first review the
structure of the bare Si(001) surface. It is well known that, on the
Si(001) surface, neighboring Si atoms pair up to form dimers \citep{ramstad1995theoretical,paz2001electron},
and we find that dimerization lowers the energy by $1.45$ eV per
dimer. The dimers can buckle (i.e., the two Si forming the dimer do
not have the same out-of-plane $z$ coordinate) which lowers their
energy. If nearby dimers buckle in opposite ways, higher order reconstructions
occur. We summarize the energies of these reconstructions in \tabref{Si_surf}
(we refer the reader to the cited works for detailed descriptions
of these surface configurations). There is a strong drive for the
surface Si atoms to dimerize (transition from a $1\times1$ to a $2\times1$
unit cell) and a weaker energetic drive to organize the dimers into
structures with periodicities larger than $2\times1$. Because the
metastable configurations of the ZrO$_{2}$ monolayers we found above
have unit cells that are $2\times1$ or smaller, we have limited our
search for Si/ZrO$_{2}$ interfaces to $2\times1$ simulation cells.

\begin{table}
\begin{centering}
\begin{tabular}{cccc}
\toprule 
\addlinespace[0.3cm]
Si surface & Energy (eV/dimer)  & Ref. \citep{ramstad1995theoretical} & Ref. \citep{paz2001electron}\tabularnewline\addlinespace[0.3cm]
\midrule
\addlinespace[0.1cm]
\midrule 
flat $p(2\times1)$ & $\equiv0.00$ & $\equiv0.00$ & $\equiv0.00$\tabularnewline
\midrule 
buckled $p(2\times1)$ & $-0.20$ & $-0.12$ & $-0.13$\tabularnewline
\midrule 
buckled $p(2\times2)$  & $-0.28$ & $-0.17$ & $-0.23$\tabularnewline
\midrule 
buckled $c(4\times2)$ & $-0.27$ & $-0.17$ & $-0.24$\tabularnewline
\bottomrule
\end{tabular}
\par\end{centering}
\caption[Energies of the lowest energy Si(001) surface reconstructions.]{\label{tab:Si_surf}Energies of the lowest energy Si(001) surface
reconstructions per dimer. Two theoretical references are presented
alongside our computed results. See the cited works for details of
the listed reconstructions.}
\end{table}

\subsubsection{Structure of the monolayers on silicon}

We have searched the configuration space for ZrO$_{2}$ on Si(001)
as follows: First, we have created a $3\times3\times2$ grid of points
inside the $2\times1$ in-plane unit cell on top of the bare Si surface
where a Zr atom is placed (the $3\times3$ grid corresponds to points
in the $xy$-plane and the $\times2$ corresponds to the vertical
distance from the substrate). A flat and high symmetry $1\times1$
zirconia monolayer is generated such that it includes this Zr atom.
For each such structure, the atoms in the Si surface layer and the
ZrO$_{2}$ monolayer are randomly and slightly displaced to generate
$5$ initial positions. This procedure, which yields $3\times3\times2\times5=90$
configurations, is done for dimerized and non-dimerized Si surfaces,
so that there are $180$ initial configurations in total. We have
then relaxed all the atoms in ZrO$_{2}$ and the top 4 layers of silicon
substrate to find local energy minima.

\begin{figure}
\begin{centering}
\includegraphics[width=1\columnwidth]{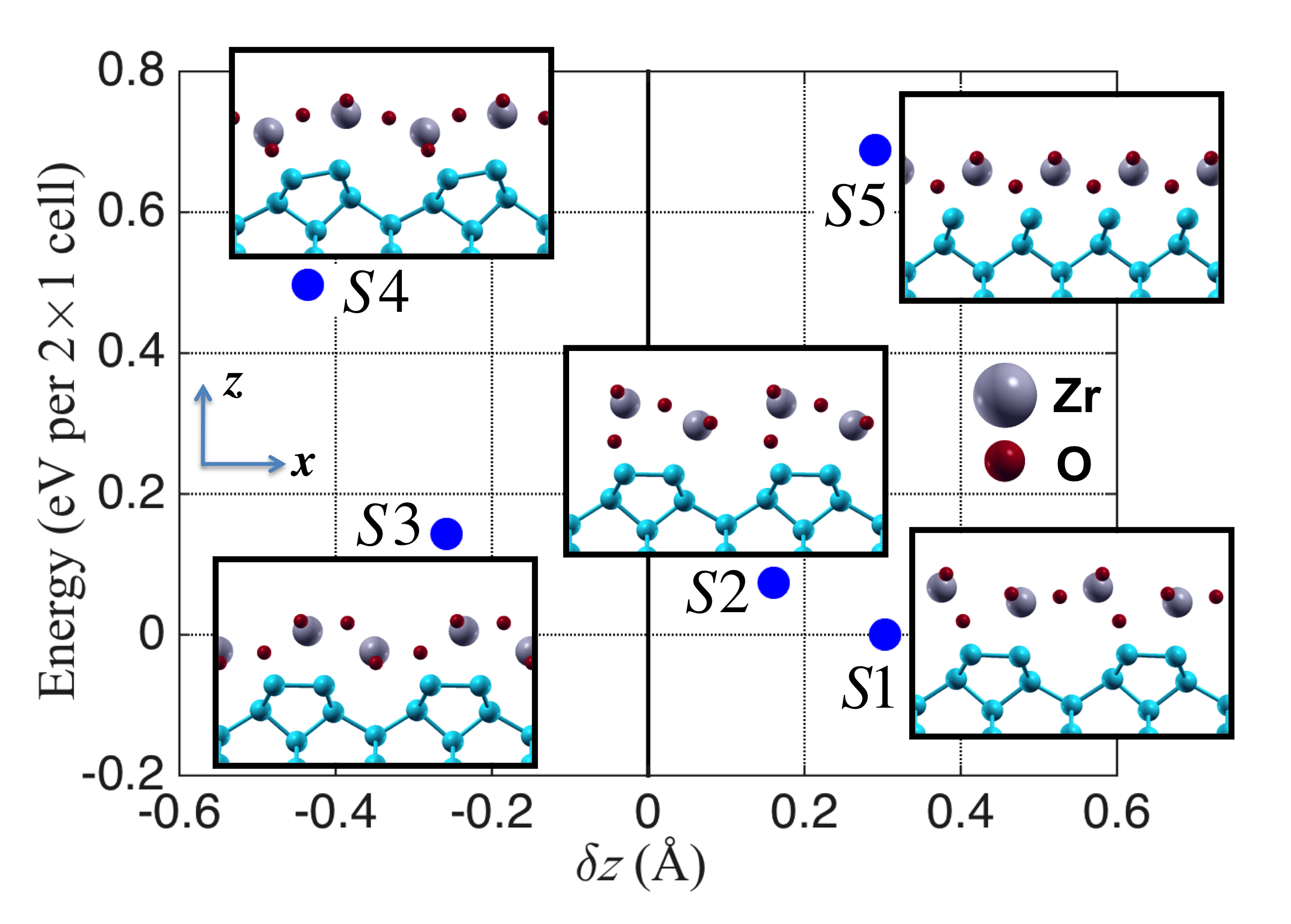}
\par\end{centering}
\caption[Five lowest energy configurations of ZrO$_{2}$ monolayers on Si.]{\label{fig:SiZrO2_en}Five lowest energy configurations of ZrO$_{2}$
monolayers on Si. $\delta z\equiv\overline{z\left(\text{Zr}\right)}-\overline{z\left(\text{O}\right)}$
is a measure of ionic out-of-plane polarization for the monolayers.
Energies are listed in eV per $2\times1$ in-plane cell measured with
respect to the lowest energy structure $S1$.}
\end{figure}

We present the five lowest energy structures we have obtained in \figref{SiZrO2_en}.
The horizontal axis is a quantity that describes the ionic polarization
of the ZrO$_{2}$ monolayer and is defined as the mean vertical Zr-O
separation $\delta z\equiv\overline{z\left(\text{Zr}\right)}-\overline{z\left(\text{O}\right)}$,
where over-bars mean averaging of the coordinate over the atoms of
that type in the structure. The vertical axis is the energy in eV
per $2\times1$ cell measured with respect to the lowest energy structure,
labeled $S1$. The energies of $S1$ through $S5$ are also listed
in \tabref{SiZrO2_en}.

\begin{table}
\begin{centering}
\begin{tabular}{cccccc}
\toprule 
\addlinespace[0.3cm]
 & $\ \ \ S1\ \ \ $ & $\ \ \ S2\ \ \ $ & $\ \ \ S3\ \ \ $ & $\ \ \ S4\ \ \ $ & $\ \ \ S5\ \ \ $\tabularnewline\addlinespace[0.3cm]
\midrule
\addlinespace[0.1cm]
\midrule 
Energy & \multirow{2}{*}{$\equiv0.00$} & \multirow{2}{*}{0.07} & \multirow{2}{*}{0.14} & \multirow{2}{*}{0.50} & \multirow{2}{*}{0.69}\tabularnewline
(eV per $2\times1$ cell) &  &  &  &  & \tabularnewline
\bottomrule
\end{tabular}
\par\end{centering}
\caption[Energies of the five lowest energy configurations of ZrO$_{2}$ monolayers
on Si.]{\label{tab:SiZrO2_en}Energies of the five lowest energy configurations
of ZrO$_{2}$ monolayers on Si as labeled in \figref{SiZrO2_en}.}
\end{table}

First, the metastable configurations lie on both sides of the $\delta z=0$
line, which means that there is no polarization direction that is
strongly preferred. Second, we find that the four lowest energy structures
have a $2\times1$ periodicity with intact Si dimers. (In addition
to $S5$, we have found three more $1\times1$ structures with broken
dimers at energies higher than 1 eV that are not shown.) The energy
difference of $0.69$ eV per dimer between the lowest energy $1\times1$
and the lowest energy $2\times1$ structures (i.e. $S5$ and $S1$)
is half of the energy of dimerization on the bare Si surface. Moreover,
the length of the dimer in $S1$ is $2.42\ \text{\AA}$ which is longer
than the $2.31\ \text{\AA}$ on the bare surface. Therefore, in general,
the Si dimers are weakened but not broken by the ZrO$_{2}$ monolayer
for the more stable low-energy structures.

Third, we notice that for each configuration shown in \figref{SiZrO2_en},
a physically equivalent configuration is obtained by a mirror reflection
by the $yz-$plane, which doubles the number of metastable structures
in the configuration space. For our analysis of transitions between
these configurations, we make the reasonable assumption that silicon
dimers remain intact during the transition between two dimerized configurations.
Hence, we reflect the atomic positions through a $yz$-plane which
keeps the dimers in place in order to obtain the geometrically inequivalent
(but physically identical) set of structures $\overline{S1}$, $\overline{S2}$
etc.

\subsubsection{Transitions between low energy states}

We have computed the minimum energy transition paths between the three
lowest energy configurations and their symmetry related counterparts
($S1,\overline{S1},S2,\overline{S2},S3,\overline{S3}$). When applying
the NEB method to find transition states, each atom in the initial
configuration is mapped to an atom in the final configuration. In
principle, all possible matching choices should be attempted in order
to find all inequivalent transition paths and energy barriers. However,
this is neither practical nor physically necessary. For the case of
free standing ZrO$_{2}$, in all the minimum energy configurations,
all atomic $(x,y)$ coordinates line on a square grid, and by making
the reasonable assumption that atoms do not swap sites during the
transition, we can dramatically reduce the number of possible transition
paths under consideration. Hence, we matched each atom in the initial
configuration with the atom that sits at the same $(x,y)$ site in
the final configuration in order to perform the NEB calculations.
Even though no fixed square grid exists for the ZrO$_{2}$/Si case
that applies to all the configurations, similar considerations are
possible: (1) For the six configurations of interest, both Zr atoms
and two out of the four O atoms in a unit cell align along the $y$-direction
with the Si dimers ($y=0.5a_{\text{lat}}$), and the other two O atoms
lie half way between consecutive dimers ($y=0$). Both along the $x$-
and the $y$-directions, atomic chains of $\ldots$-Zr-O-Zr-O-$\ldots$
exist in all cases. So for each configuration, we can make a square
grid in the $xy-$plane such that one Zr per cell sits at a lattice
site and the other atoms are very close to the other lattice sites.
For each transition process, the grid is assumed only to shift in
the $x$-direction. (2) Because of the high energy cost of breaking
Si dimers on the bare Si(001) surface, we assume that the dimers remain
intact during a transition. (3) We assume that $\ldots$-Zr-O-Zr-O-$\ldots$
chains along the $y$-direction remain intact during a transition,
so no movement in the $y$-direction is considered.

\begin{figure}
\begin{centering}
\includegraphics[width=0.85\columnwidth]{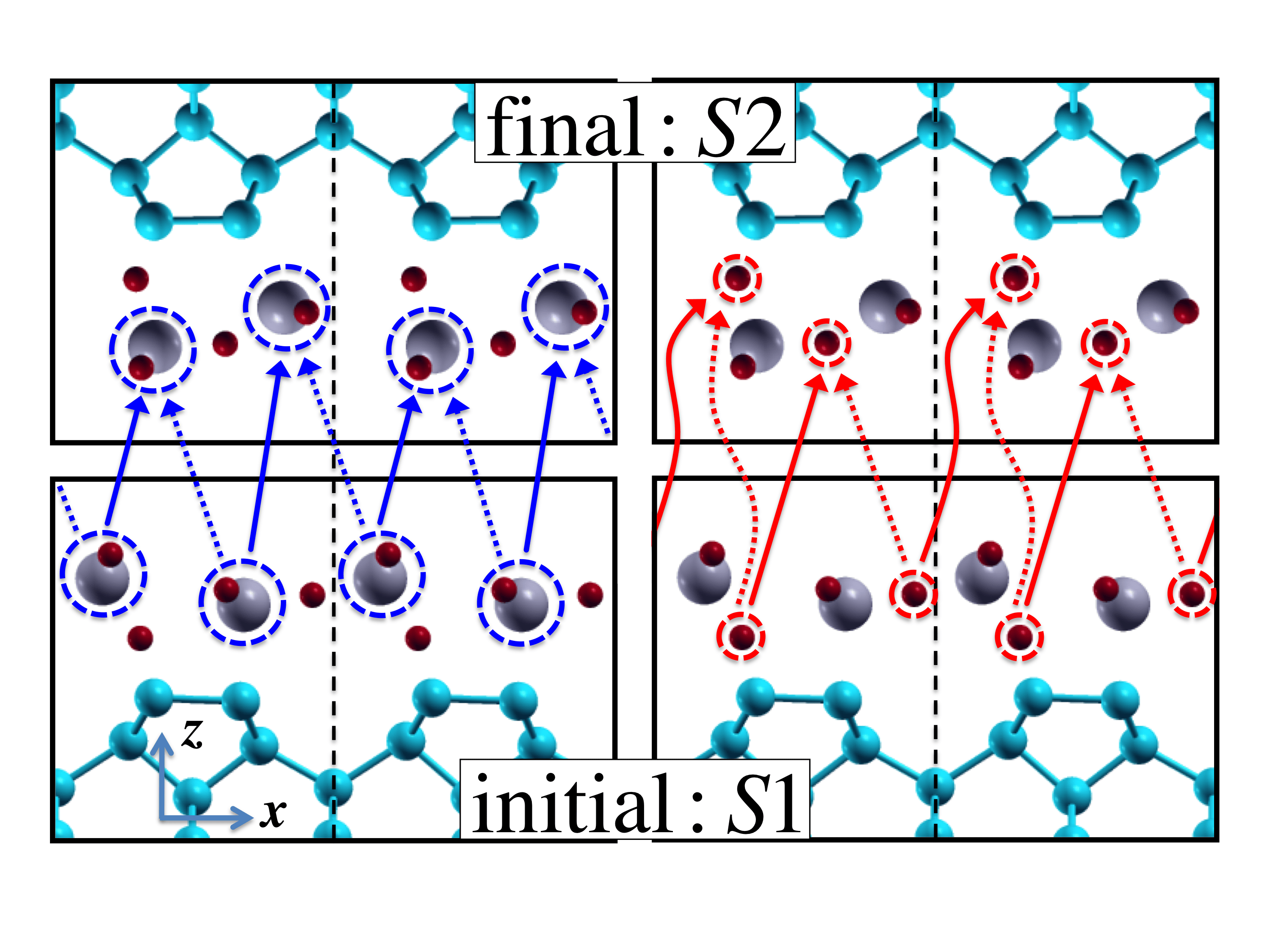}
\par\end{centering}
\caption[The possible matchings for the $S1\rightarrow S2$ transition for
the NEB simulation.]{\label{fig:SiZrO2_match}The possible matchings for the $S1\rightarrow S2$
transition for the NEB simulation. The $S2$ structure is displayed
upside down to allow for ease of understanding the matching. In the
left panels, two possible choices for the two Zr-O pairs (or chains)
in the $S1$ unit cell that are to be matched to the Zr-O pairs (or
chains) in the structure $S2$ are shown. The set of solid arrows
corresponds to one choice, and the set of dotted lines corresponds
to another choice. Similarly, two choices for the remaining oxygens
are displayed in the right panels. See text for further details of
the described matchings. Two periodic copies of $2\times1$ cells
are are shown in each case, and a dashed line is drawn to separate
the copies. }
\end{figure}

By using these constraints, we can reduce the number of possible matchings
to four for each transition. We demonstrate these choices for the
transition $S1\rightarrow S2$ in \figref{SiZrO2_match}. The final
state $S2$ is displayed upside down in order allow for a clearer
illustration of atomic matchings. In the left panel, $\ldots$-Zr-O-Zr-O-$\ldots$
chains along the $y$-direction are circled by blue dashed rings.
There are two possible ways in which the chains in $S1$ can be matched
to the chains in $S2$ that do not cause large scale rearrangements.
One of these matchings is shown as solid arrows, and the other is
shown as dotted arrows. In the right panel, the same exercise is repeated
for the remaining oxygens (circled by red dashed rings). Therefore
there are $2\times2=4$ matchings in total. Note that the reverse
processes correspond to the set of matchings that obey our rules for
the transition $S2\rightarrow S1$.

The resulting smallest energy barriers are listed in \tabref{SiZrO2_neb}.
Notice that the nine listed transitions cover all the possible transitions
because, e.g., the transition $S1\leftrightarrow\overline{S2}$ is
related by symmetry to $\overline{S1}\leftrightarrow S2$. We observe
that the transitions within the set of unbarred states are about 1
eV smaller than the transitions between unbarred and barred states.
This is understood as follows: for all six structures, there is one
oxygen per cell which binds to a silicon atom. The transitions that
leave that oxygen in place (such as the dotted arrows in the right
panels of \figref{SiZrO2_match}) have lower energy barriers. A transition
between an unbarred state and a barred state necessarily involves
displacing that oxygen and breaking the strong Si-O bond. Therefore
a low energy path is not possible in such a case.

\begin{table}
\begin{centering}
\begin{tabular}{ccc}
\toprule 
\addlinespace[0.3cm]
Transition & $\ \ E_{\text{barrier}}\left(\rightarrow\right)$ (eV) & $\ \ E_{\text{barrier}}\left(\leftarrow\right)$ (eV)\tabularnewline\addlinespace[0.3cm]
\midrule
\addlinespace[0.1cm]
\midrule 
$S1\leftrightarrow\overline{S1}$ & 1.63 & 1.63\tabularnewline
\midrule 
$S1\leftrightarrow S2$ & 0.79 & 0.71\tabularnewline
\midrule 
$S1\leftrightarrow\overline{S2}$ & 1.60 & 1.52\tabularnewline
\midrule 
$S1\leftrightarrow S3$ & 0.79 & 0.65\tabularnewline
\midrule 
$S1\leftrightarrow\overline{S3}$ & 1.60 & 1.46\tabularnewline
\midrule 
$S2\leftrightarrow\overline{S2}$ & 2.48 & 2.48\tabularnewline
\midrule 
$S2\leftrightarrow S3$ & 0.23 & 0.17\tabularnewline
\midrule 
$S2\leftrightarrow\overline{S3}$ & 1.57 & 1.51\tabularnewline
\midrule 
$S3\leftrightarrow\overline{S3}$ & 1.77 & 1.77\tabularnewline
\bottomrule
\end{tabular}
\par\end{centering}
\caption[Transition barriers, calculated via the NEB method, between pairs
of low energy configurations of ZrO$_{2}$ monolayers on Si(001).]{\label{tab:SiZrO2_neb}Transition barriers, calculated via the NEB
method, between pairs of low energy configurations of ZrO$_{2}$ monolayers
on Si(001). Energy barriers are reported in eV per $2\times1$ cell.
The central and rightmost columns show the barriers going in both
directions (as indicated by the arrow directions).}
\end{table}

\begin{figure}
\begin{centering}
\includegraphics[width=1\columnwidth]{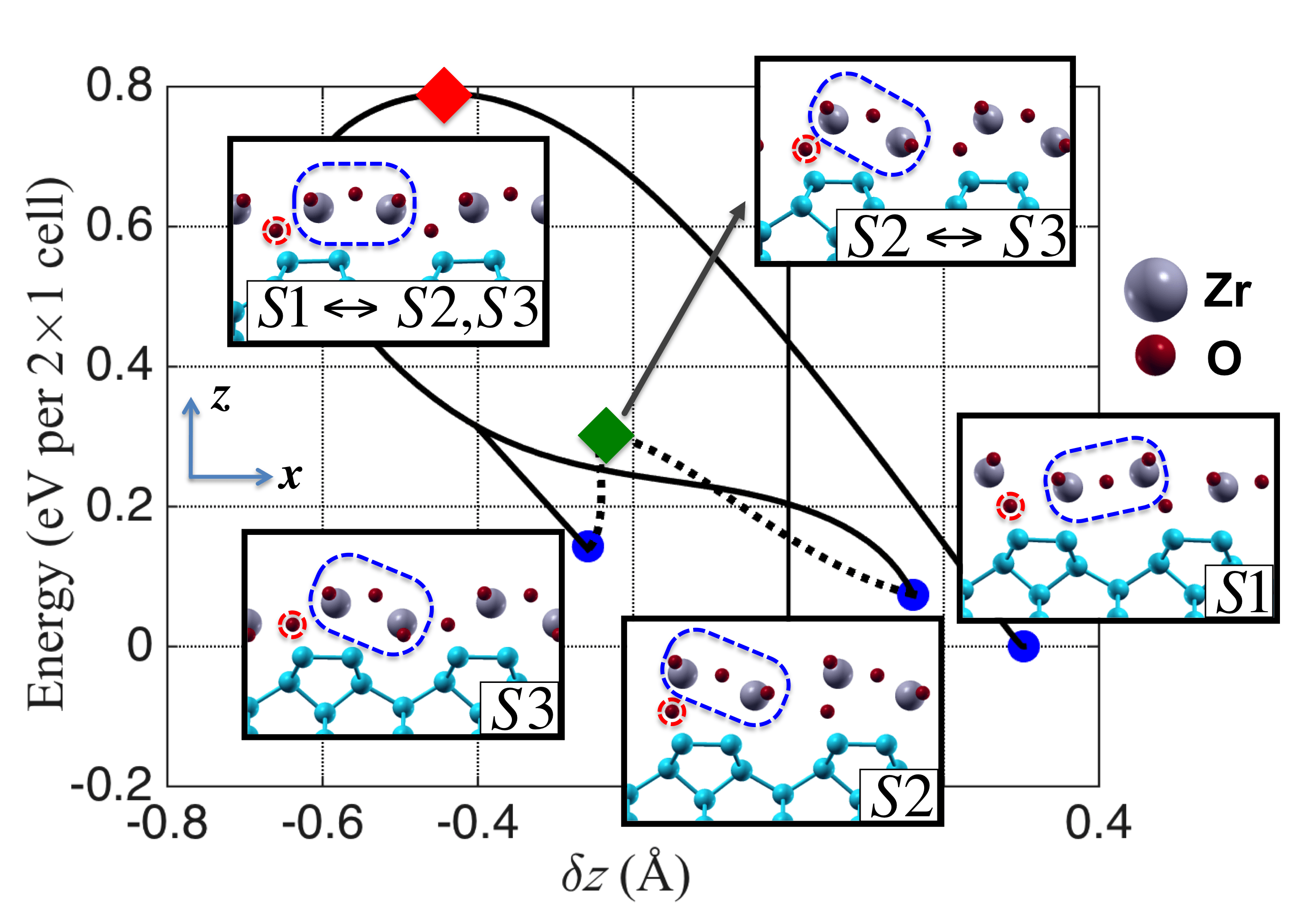}
\par\end{centering}
\caption[Three lowest energy configurations of ZrO$_{2}$ monolayers on Si
and the transition paths between them calculated via the NEB method.]{\label{fig:SiZrO2_NEB}Three lowest energy configurations of ZrO$_{2}$
monolayers on Si and the transition paths between them calculated
via the NEB method. The solid curve corresponds to the transitions
$S1\leftrightarrow\left(S2,S3\right)$ that share a transition state
denoted by a red diamond. The dotted curve corresponds to the transition
$S2\leftrightarrow S3$ which has a transition state denoted by a
green diamond. The circled oxygen atoms remain in place during the
transitions, and the circled groups of five atoms move as a block
with small internal displacements.}
\end{figure}

Focusing on the three low energy transitions, i.e. $S1\leftrightarrow S2$,
$S1\leftrightarrow S3$ and $S2\leftrightarrow S3$, we plot energy
vs $\delta z$ curves in \figref{SiZrO2_NEB}. The transition state
of $S2\leftrightarrow S3$ (dotted curve) and the shared transition
state of $S1\leftrightarrow S2$ and $S1\leftrightarrow S3$ (solid
curves) are marked by diamonds on the plot and their configurations
are displayed. During these transitions, the oxygen atom that is bonded
to a silicon (circled by red dashed rings in the figure) remains in
place, while the remaining 5 atoms in the ZrO$_{2}$ layer (inside
the blue dashed rounded rectangles) move in concert. Because this
movement does not significantly alter the chemistry of the interface,
the energy barriers are relatively low.

Because of the rich landscape of stable configurations at low energy
with similar chemical bonding and small structural differences, we
predict that growing large single-crystalline epitaxial films of ZrO$_{2}$
on Si(001) should be challenging. However, epitaxy may not be a necessary
condition for ferroelectricity in this system. A close examination
of the structures shown in \figref{SiZrO2_NEB} indicates that the
symmetry of the silicon surface, as well as the inherently rumpled
structure of ZrO$_{2}$, give rise to the switchable polarization.
The switching of the dipole occurs by a continuous displacement of
a group of atoms in the unit cell, while one oxygen remains in place.
No significant chemical change occurs during these transitions. We
note that open channels in the dimerized (001) face of silicon allow
for the motion of the oxide atoms lacking silicon nearest neighbors,
which stabilizes the three low-energy polar ZrO$_{2}$ structures.

\subsubsection{Coupling of polarization to electronic states in Si}

\begin{figure}
\begin{centering}
\includegraphics[width=0.9\columnwidth]{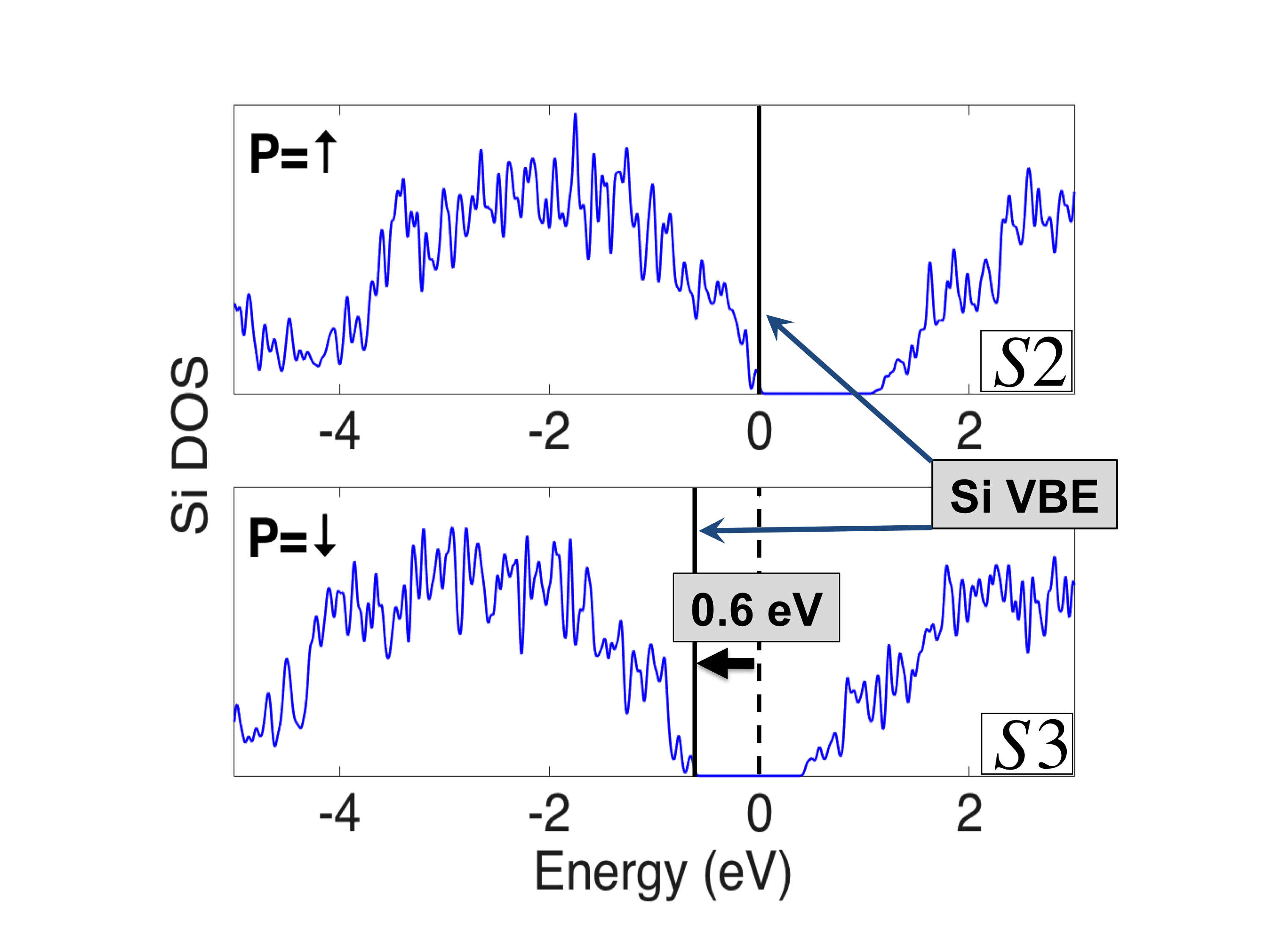}
\par\end{centering}
\caption[Density of states in an interior Si layer with the ZrO$_{2}$ film
in its upwardly polarized and downwardly polarized forms.]{\label{fig:SiZrO2_DOS}Density of states in an interior Si layer
with the ZrO$_{2}$ film in its upwardly polarized ($S2$) and downwardly
polarized ($S3$) forms. There exists a valence band edge (VBE) shift
between the \textquotedbl up\textquotedbl{} state (top) and the \textquotedbl down\textquotedbl{}
state (bottom). This figure is reproduced from Ref. \citep{dogan2018singleatomic}.}
\end{figure}

In addition to the prediction that the three lowest energy structures
may coexist in monolayer form, in \subsecref{Domain} we will explain
why, at temperatures of practical interest, structures $S2$ and $S3$
should be the dominant motifs in the monolayer structure. Because
of the large difference in polarization together with a low energy
barrier between these two structures, we believe that the polarization
switching described in Ref. \citep{dogan2018singleatomic} should
correspond to switching between $S2$ and $S3$. A first and simple
corroboration involves showing that the change in the silicon Fermi
level observed in the experiment is comparable with our theoretical
prediction. In \figref{SiZrO2_DOS}, we plot the density of states
(DOS) of the ZrO$_{2}$/Si system projected onto an interior layer
of the Si substrate for the cases of interface structures $S2$ and
$S3$. We set the energy of the Si valence band edge (VBE) of $S2$
to zero and align the vacuum energy level in $S3$ to the vacuum energy
energy in $S2$. We find a $0.6$ eV VBE shift in Si, which is somewhat
larger than, but comparable to, the experimental value of $0.4$ eV.
We believe that this is due to the fact that the experimental monolayers
are not epitaxial but amorphous with multiple structural motifs present,
so that application of the electric field is not as effective at polarization
switching as is assumed in our clean, epitaxial and ordered theoretical
simulations.

\subsection{Domain energetics\label{subsec:Domain}}

Up to this point, our theoretical study of the ZrO$_{2}$ monolayers
on the Si(001) surface has shown that (meta)stable configurations
with varying polarizations are present. We have also demonstrated
that transitions between some of the lowest energy configurations
do not require complicated rearrangements of atoms and have low energy
barriers. Because of these findings, as well as the fact that the
experimental monolayer is amorphous, we expect there to be a multi-domain
character to these monolayers at or near room temperature ($k_{\text{B}}T=0.026$).
However, directly calculating the energy of a multi-domain region
of the system for an area larger than a few primitive unit cells is
not feasible. In this section, we describe an approximate model Hamiltonian
method to compute the energies of arbitrary regions of multiple domains,
and use Monte Carlo simulations to find thermodynamic ground states
at finite temperatures. 

\subsubsection{Domain wall energies}

In order to investigate the behavior of finite domains, we have developed
a lattice model where every $2\times1$ in-plane cell is treated as
a site in a two dimensional lattice which couples to its neighbors
via an interaction energy. Similar models have been proposed for other
two dimensional systems \citep{bune1998twodimensional}. Such a model
is reasonable if the interface (domain wall) between domains of different
states is sharp, i.e., the atomic positions a few unit cells away
from a domain boundary are indistinguishable from the atomic positions
in the center of the domain. To find the degree of locality and the
energy costs of the domain walls, we have computed domain wall energies
as a function of domain size.

\begin{figure}
\begin{centering}
\includegraphics[width=1\columnwidth]{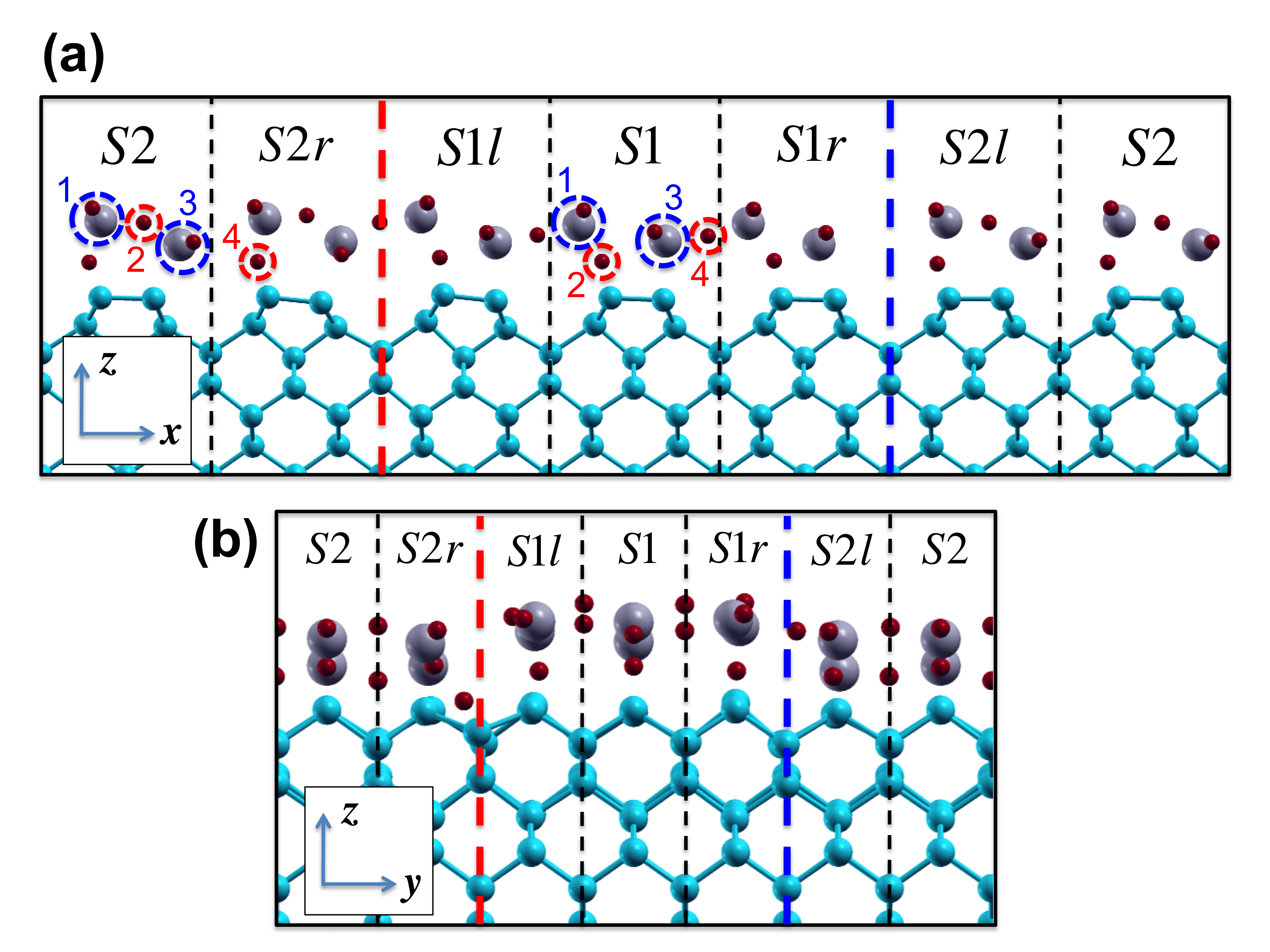}
\par\end{centering}
\caption[Simulation arrangements to compute the domain boundary energies between
$S1$ and $S2$.]{\label{fig:SiZrO2_latt_cell}Simulation arrangements to compute the
domain boundary energies between $S1$ and $S2$. (a) 3 cells each
of $S1$ and $S2$ are stacked along the $x$-direction to form straight
domain boundaries in the $y$-direction. The numberings of atomic
groups within the unit cells are displayed using dashed circles. The
boundary on the right (blue) is initially built by the atomic groups
1, 2 and 3 from $S1$ and $4$ from $S2$ in the unit cell to the
left of the boundary (labelled $S1r$), and the atomic groups 1, 2,
3 and 4 from $S2$ in the unit cell to the right of the boundary (labelled
$S2l$). The boundary on the left (red) is constructed to preserve
the number of atomic groups from each cell. (b) 3 cells each of $S1$
and $S2$ are stacked along the $y$-direction to form straight domain
boundaries in the $x$-direction. Fully relaxed boundary configurations
are shown.}
\end{figure}

Sample simulation arrangements are shown in \figref{SiZrO2_latt_cell}.
In (a) and (b), domain walls along the $y$- and $x$-directions are
formed, respectively, between the configurations $S1$ and $S2$.
Three unit cells of $S1$ and $S2$ each are generated and attached
together to build larger simulation cells to model the domain walls:
$12\times1$ and $2\times6$ cells to simulate the domain boundaries
along the $y$- and $x$-directions, respectively. In each of the
3 unit wide domains, the center unit is fixed to the atomic configuration
of the corresponding uniform system. In \figref{SiZrO2_latt_cell},
for the $S1$ domain, the atoms in the unit labelled $S1$ are fixed,
and the atoms in the units $S1l$ and $S1r$ are relaxed. The same
is true for $S2$, but for clarity, fixed units of $S2$ are displayed
on both sides. We then compute the domain wall energy between $S1$
and $S2$ by subtracting $3E\left(S1\right)+3E\left(S2\right)$ from
the total energy of this supercell and dividing by two. We have checked
for a few test cases that increasing the domain width from 3 to 5
cells changes the domain wall energies by small amounts on the order
of 1-10 meV while typical domain wall energies are larger than 100
meV (see \tabref{SiZrO2_latt_J}). This, together with visualization
of the resulting structures, convinces us that the domains are sufficiently
local for us to treat the domain walls as being sharp. Note that there
are two inequivalent boundaries between $S1$ and $S2$ along a given
direction. In \figref{SiZrO2_latt_cell}, these boundaries are shown
as red and blue dashed lines. Due to the periodicity of simulation
cells, it is not possible to compute the energies of these two boundaries
independently, so we are forced to assume that their energies are
equal.

The final step in determining the domain boundary energies is to survey
the configuration space available for a given boundary. For that purpose,
for each domain boundary we have generated a number of initial configurations
depending on the direction of the boundary:
\begin{itemize}
\item For a boundary along the $y$-direction such as in \figref{SiZrO2_latt_cell}(a),
we have generated five initial configurations as follows. For each
domain state (e.g., $S1$ or $S2$), we have labeled the Zr-O pairs
along the $y$-direction and the remaining oxygens and numbered them
in an increasing order in the $x$-direction. In the figure, the labelling
for states $S1$ and $S2$ is shown. Note that for each cell, the
sequence starts with a Zr-O pair and ends with an O atom. Hence in
some cases the oxygen labelled 4 lies beyond the unit cell to which
it belongs, such as in $S2$. To build a domain boundary such as the
$S1r$-$S2l$ (shown as a blue dashed line), we first place the atomic
groups numbered $1-4$ from $S1$ to the left hand side of the boundary,
and the atomic groups numbered $1-4$ from $S2$ to the right hand
side of the boundary. This constitutes our first initial configuration.
The second configuration is obtained by replacing atom $4$ from $S1$
on the left hand side by atom $4$ from $S2$. The third is obtained
by replacing both group $3$ and atom $4$ from $S1$ by $3$ and
$4$ from $S2$. The fourth choice is to replace atomic group $1$
from $S2$ on the right hand side by group $1$ from $S1$; and, lastly,
the fifth choice is to replace $1$ and $2$ from $S2$ by $1$ and
$2$ from $S1$. The opposite operation is performed at the other
boundary such as $S2r$-$S1l$ (shown as a red dashed line). We then
take the smallest of the five computed domain energies as the final
energy. Note that the relaxed structure shown in the \figref{SiZrO2_latt_cell}(a)
for the $S1$-$S2$ domain boundaries is obtained via choice $\#2$
for the $S1r$-$S2l$ boundary.
\item For a boundary along the $x$-direction such as in \figref{SiZrO2_latt_cell}(b),
we have generated a few initial configurations by slightly and randomly
displacing the two oxygen atoms at the boundary along the $y$-direction
in order to break the $y\rightarrow-y$ symmetry inherent to these
structures.
\end{itemize}

\subsubsection{Construction of a lattice model}

\begin{figure}
\begin{centering}
\includegraphics[width=0.9\columnwidth]{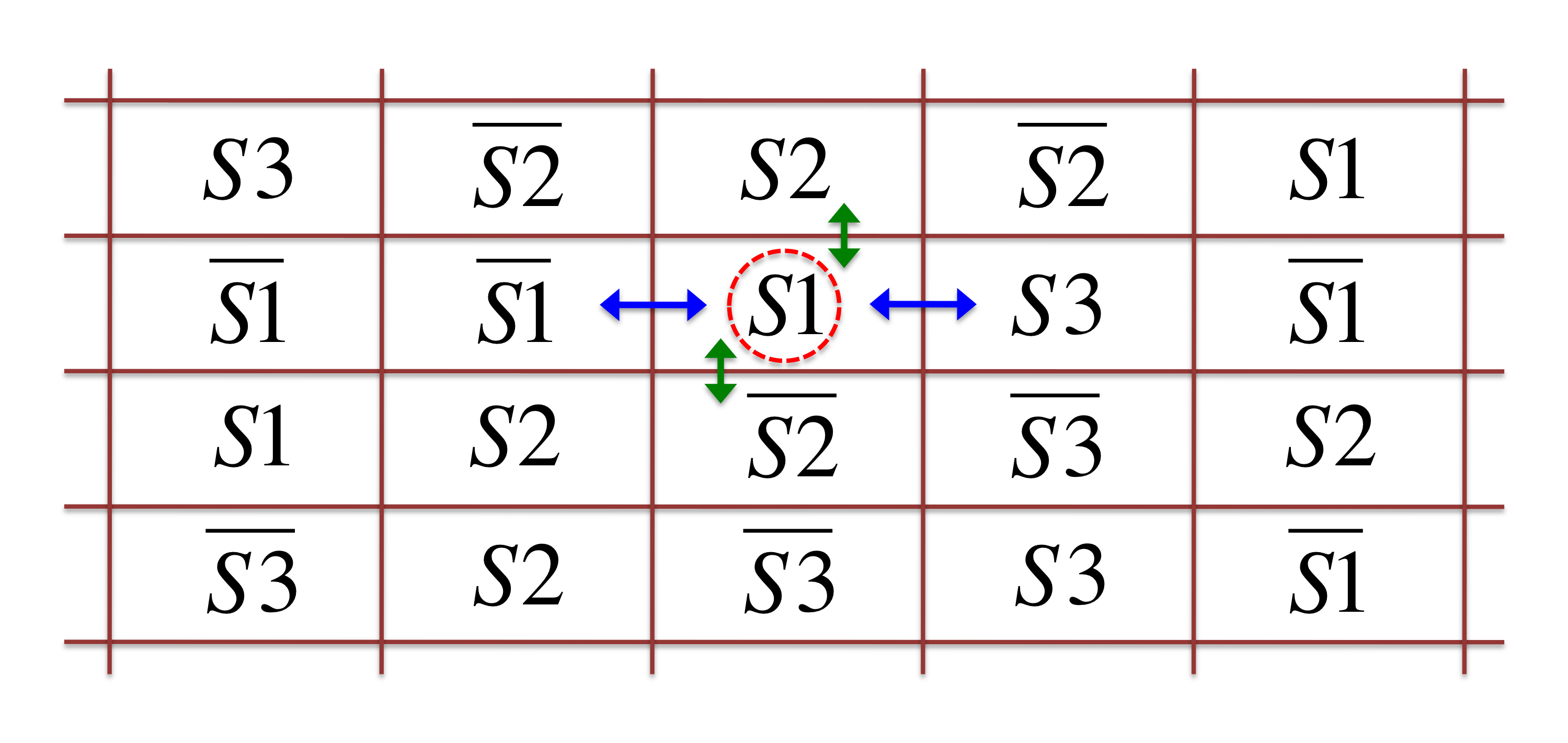}
\par\end{centering}
\caption[An example configuration of the two dimensional lattice that approximates
the ZrO$_{2}$ monolayer on Si as a multi-domain system.]{\label{fig:SiZrO2_lattice}An example configuration of the two dimensional
lattice that approximates the ZrO$_{2}$ monolayer on Si as a multi-domain
system. Nearest neighbor sites couple through the coefficients $J_{x}$
(blue arrows) and $J_{y}$ (green arrows).}
\end{figure}

Once we have the library of domain boundary energies for every pair
of states along the $x$- and $y$-directions described above, we
approximate the energy of the system with an arbitrary configuration
of domains by a two-dimensional anisotropic lattice Hamiltonian on
a square lattice:

\begin{eqnarray}
H & = & \sum_{i,j}E\left(\sigma\left(i,j\right)\right)+\sum_{i,j}J_{x}\left(\sigma\left(i,j\right),\sigma\left(i+1,j\right)\right)\nonumber \\
 &  & +\sum_{i,j}J_{y}\left(\sigma\left(i,j\right),\sigma\left(i,j+1\right)\right),
\end{eqnarray}
where $\sigma\left(i\right)$ donates the state at a given site $i$,
$E\left(\sigma\left(i\right)\right)$ is the energy (per $2\times1$
unit cell) of state $\sigma\left(i\right)$ for a uniform system in
that state, and $J_{\alpha}\left(\sigma\left(i\right),\sigma\left(j\right)\right)$
is the energy of interaction (i.e., domain wall energy) between the
neighboring states $i,j$ in the axial direction $\alpha$. In our
model, only nearest neighbor interactions are included. Because of
the anisotropic nature of the film (the $x$- and $y$-directions
are fundamentally different due to Si dimerization), the interaction
term must distinguish between directions $x$ and $y$ so that $J_{x}$
and $J_{y}$ differ. The domain boundary energies calculated via DFT
simulations are employed as nearest neighbor interaction energies
in this model. In \figref{SiZrO2_lattice}, we illustrate an arbitrary
configuration of such a lattice. As an example, the state $S1$ in
the middle column couples to $\overline{S1}$ and $S3$ via $J_{x}\left(S1,\overline{S1}\right)$
and $J_{x}\left(S1,S3\right)$, respectively, and to $S2$ and $\overline{S2}$
via $J_{y}\left(S1,S2\right)$ and $J_{y}\left(S1,\overline{S2}\right)$,
respectively.

For a model with $N$ distinct states, our interaction matrices $J_{\alpha}$
($\alpha=x,y$) have the following properties:
\begin{itemize}
\item The interaction energy between the sites of the same kind is zero
by definition, $J_{\alpha}\left(\sigma_{i},\sigma_{i}\right)=0$.
Hence the number of non-zero entries is $N^{2}-N$.
\item We have assumed that the domain wall energy between states $\sigma_{i}$
and $\sigma_{j}$ remains the same if we swap the states. Therefore
the interaction matrices are symmetric $J_{\alpha}\left(\sigma_{i},\sigma_{j}\right)=J_{\alpha}\left(\sigma_{j},\sigma_{i}\right)$,
reducing the number of unique non-zero entries to $\frac{1}{2}\left(N^{2}-N\right)$.
\item In our particular system, every state has a counterpart which is obtained
by the reflection $x\rightarrow-x$. Hence, e.g., the domain wall
between $\overline{S1}$ and $\overline{S2}$ can be obtained from
the domain wall between $S1$ and $S2$ by applying a single symmetry
operation. Therefore many of the entires of $J_{\alpha}\left(\sigma_{i},\sigma_{j}\right)$
are paired up in this way which further reduces the number of unique
entries further to $\frac{1}{4}N^{2}$.
\end{itemize}
In \tabref{SiZrO2_latt_J}, we list the unique entries of $J_{\alpha}\left(\sigma_{i},\sigma_{j}\right)$
for states $\sigma$ ranging over the the six lowest energy states.
Note that since $N=6$ for this table, there are $\frac{1}{4}6^{2}=9$
entries in the table. Because the unit cell is $2\times1$, the couplings
$J_{x}$ are expected to be smaller than the couplings $J_{y}$, which
is generally correct. We have computed the domain wall energies for
more possible of states including $S4$, $\overline{S4}$, $S5$ and
$\overline{S5}$, and the longer list of resulting domain wall energies
(see Supplementary Material) are included in our treatment of the
lattice model below.

\begin{table}
\begin{centering}
\begin{tabular}{ccc}
\toprule 
\addlinespace[0.3cm]
Domain boundary & $\ \ J_{x}$ (eV)$\ \ $ & $\ \ J_{y}$ (eV)$\ \ $\tabularnewline\addlinespace[0.3cm]
\midrule
\addlinespace[0.1cm]
\midrule 
$S1,\overline{S1}$ & 0.26 & 1.35\tabularnewline
\midrule 
$S1,S2$ & 0.76 & 1.13\tabularnewline
\midrule 
$S1,\overline{S2}$ & 0.96 & 0.99\tabularnewline
\midrule 
$S1,S3$ & 0.61 & 4.81\tabularnewline
\midrule 
$S1,\overline{S3}$ & 0.44 & 1.75\tabularnewline
\midrule 
$S2,\overline{S2}$ & 0.38 & 1.64\tabularnewline
\midrule 
$S2,S3$ & 0.17 & 0.98\tabularnewline
\midrule 
$S2,\overline{S3}$ & 0.01 & 0.91\tabularnewline
\midrule 
$S3,\overline{S3}$ & 0.73 & 0.002\tabularnewline
\bottomrule
\end{tabular}
\par\end{centering}
\caption[Domain boundary energies computed from first principles.]{\label{tab:SiZrO2_latt_J}Domain boundary energies between low-energy
states as computed from first principles. These energies, along with
the couplings that include the states $S4$, $\overline{S4}$, $S5$
and $\overline{S5}$ reported in Table 1 of the Supplementary Material,
serve as the couplings of nearest neighbors in our lattice model.}
\end{table}

\begin{figure}
\begin{centering}
\includegraphics[width=0.9\columnwidth]{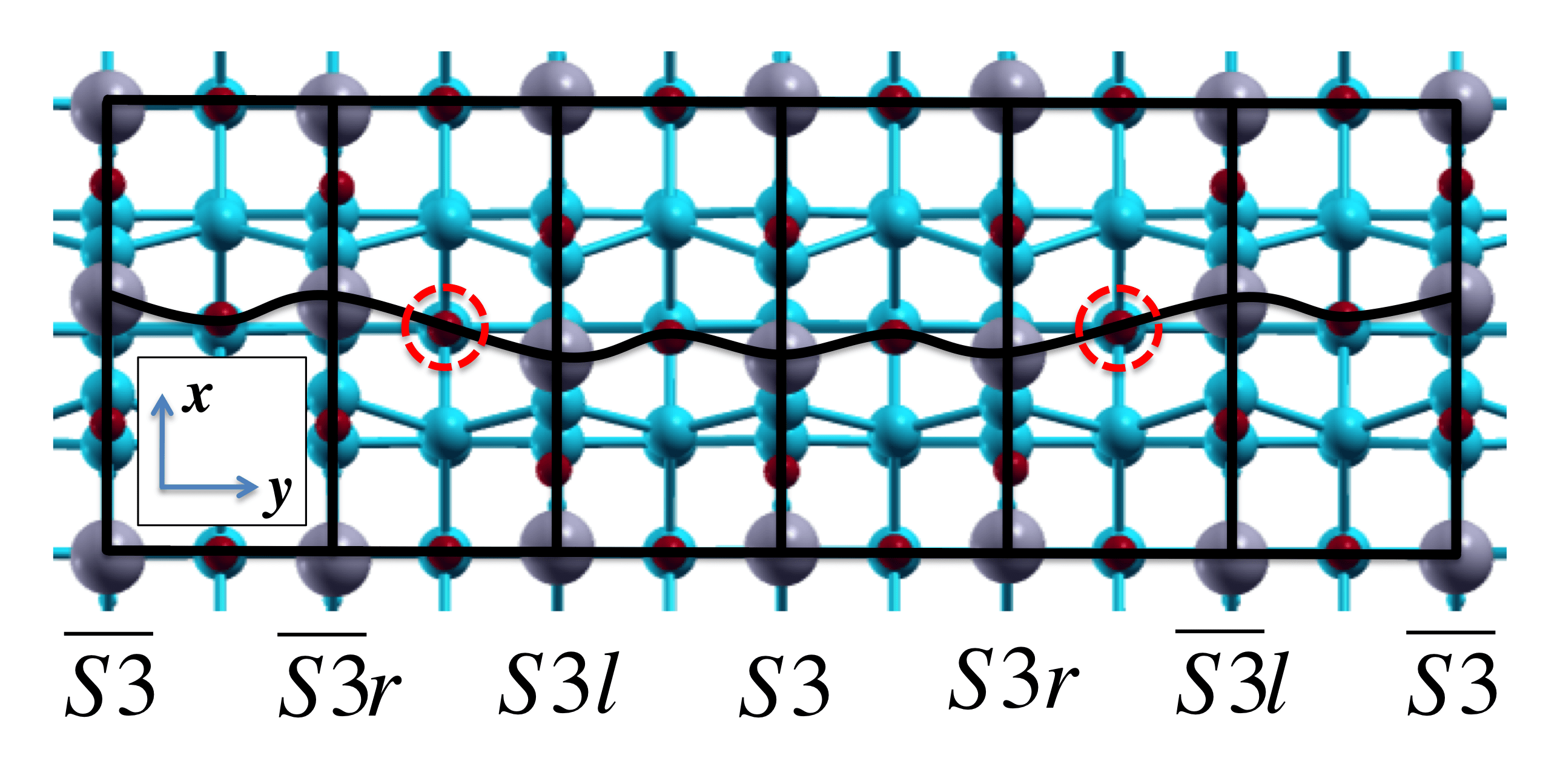}
\par\end{centering}
\caption[Top view of the domain boundaries along the $x$-direction between
$S3$ and $\overline{S3}$, computed by stacking 3 unit cells of each
structure.]{\label{fig:SiZrO2_dom_S3S3b}Top view of the domain boundaries along
the $x$-direction between $S3$ and $\overline{S3}$, computed by
stacking 3 unit cells of each structure along the $y$-direction.
The domain energy, computed to be $J_{y}\left(S3,\overline{S3}\right)=0.002\ \text{eV}$
per unit length, is very small due to the near-perfect meshing of
the $\ldots$-Zr-O-Zr-O-$\ldots$ chains in this configuration.}
\end{figure}

We notice that some of the values in \tabref{SiZrO2_latt_J}, namely
$J_{x}\left(S2,\overline{S3}\right)$ and $J_{y}\left(S3,\overline{S3}\right)$,
are very small, which is expected to be a significant factor in the
finite temperature behavior of our model. We demonstrate the domain
wall that corresponds to $J_{y}\left(S3,\overline{S3}\right)$ in
\figref{SiZrO2_dom_S3S3b} via a top view. Because one of the $\ldots$-Zr-O-Zr-O-$\ldots$
chains along the $y$-direction in the $S3$ unit cell is approximately
aligned with the valley between consecutive Si dimers along the $x$-direction,
it is approximately unchanged under the $S3\rightarrow\overline{S3}$
transformation. Therefore when $S3$ and $\overline{S3}$ cells are
attached in the $y$-direction, continuous and linear $\ldots$-Zr-O-Zr-O-$\ldots$
chains are obtained (the top and bottom black horizontal straight
lines in \figref{SiZrO2_dom_S3S3b}). The remaining $\ldots$-Zr-O-Zr-O-$\ldots$
chain in the unit cells matches imperfectly, but the distortion is
small (the winding black horizontal curve in the middle in \figref{SiZrO2_dom_S3S3b})
such that the only atom with a slightly modified environment is one
of the oxygen atoms at the domain boundary (encircled with a red dashed
ring in the figure). This near-perfect meshing of the $\ldots$-Zr-O-Zr-O-$\ldots$
chains after stacking the $S3$ and $\overline{S3}$ structures along
the $y$-direction is the cause of the very small energy cost of creating
the domain boundary.

The model we have built is a general discrete lattice model that resembles
two dimensional Ising models and, more generally, Potts models \citep{wu1982thepotts}.
However, due to the lack of any simple pattern in site energies and
couplings, it does not belong to any analytically solvable category
of models.

\subsubsection{Mean-field approach}

To understand the thermodynamic behavior of this model at finite temperature,
we begin with the standard mean-field approach which is based on the
assumption that every site interacts in an averaged manner with its
neighboring sites. For a model with $N$ states $\sigma_{1},\sigma_{2},\ldots\sigma_{N}$,
every site has a probability $p$$\left(\sigma_{i}\right)$ of being
occupied by state $\sigma_{i}$. In mean field theory, the energy
of such a site including its interactions with its nearest neighbors
is given by
\begin{eqnarray}
U\left(\sigma_{i}\right) & = & E\left(\sigma_{i}\right)+2\sum_{j=1}^{N}p\left(\sigma_{j}\right)J_{x}\left(\sigma_{i},\sigma_{j}\right)\nonumber \\
 &  & +2\sum_{j=1}^{N}p\left(\sigma_{j}\right)J_{y}\left(\sigma_{i},\sigma_{j}\right).\label{eq:U}
\end{eqnarray}

The probability $p$$\left(\sigma_{i}\right)$ is given by the the
Boltzmann factor so that
\begin{equation}
p\left(\sigma_{i}\right)=\frac{\exp\left(-\frac{U\left(\sigma_{i}\right)}{k_{\text{B}}T}\right)}{Z},
\end{equation}
where 
\begin{equation}
Z=\sum_{j=1}^{N}\exp\left(-\frac{U\left(\sigma_{j}\right)}{k_{\text{B}}T}\right)
\end{equation}
is the mean-field partition function.

\begin{figure}
\begin{centering}
\includegraphics[width=1\columnwidth]{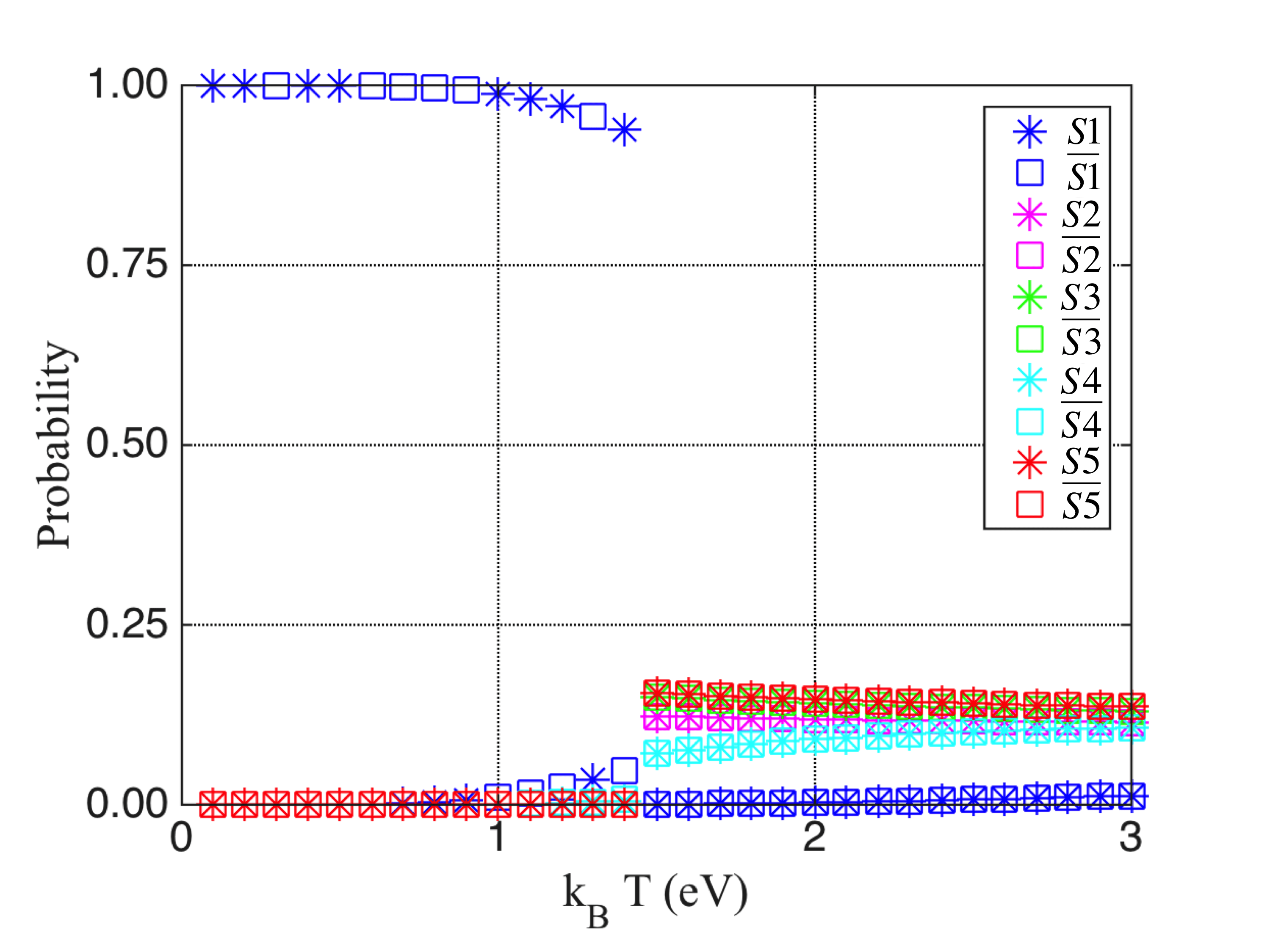}
\par\end{centering}
\caption{\label{fig:SiZrO2_latt_MF}Probabilities of finding a type of state
at an arbitrary site vs temperature, as computed by the mean-field
equations for our lattice model.}
\end{figure}

These equations form a self-consistent system of $N$ equations for
$p\left(\sigma_{i}\right)$ for a given temperature $T$ and the specified
energies $E(\sigma_{i})$ and couplings $J_{x}$, $J_{y}$. We present
the solutions of this system of equations for temperatures ranging
from 0.1 through 3.0 $eV/k_{B}$ in \figref{SiZrO2_latt_MF}. We find
that there is a first-order phase transition at a very high temperature
of $k_{\text{B}}T=1.4$ eV ($\sim$16,000 K). Below this temperature,
one of the two degenerate ground states ($S1$ or $\overline{S1})$
occupies nearly all the sites (i.e., spontaneous symmetry breaking).
Above the transition temperature, the ground states are suppressed
and the lattice gets filled by the remaining states with an approximately
equal contributions. At very high temperature (not shown in the figure),
all states have equal probability, as expected.

It is known that in simpler two dimensional lattice problems, the
mean-field approximation predicts correctly the existence of a phase
transition but overestimates the critical temperature \citep{neto2006anisotropic}.
The mean-field approach assumes that each site interacts with all
its neighbors in an uncorrelated fashion and neglects the fact that
correlation lengths are finite. Moreover, as seen from \eqref{U},
the mean-field equations sum over all neighbors and end up providing
``isotropic solutions'' (i.e., the $x$ and $y$ directions become
equivalent), which is an serious shortcoming due to the major role
anisotropy is expected to, and will, play in our system (see \ref{tab:SiZrO2_latt_J}).
In summary, we expect these mean field theory predictions to be informative
but not quantitatively accurate.

\subsubsection{Monte Carlo simulations}

For a better understanding of our model at temperatures of practical
interest, we have employed classical Monte Carlo simulations with
a modified version of the Wolff cluster algorithm \citep{swendsen1987nonuniversal,wolff1989collective}
that we have developed. For further details of the method, we refer
the reader to the Supplementary Material. We have run simulations
in a $50\times150$ lattice with free boundary conditions (i.e., the
lattice is a finite-sized system with zero couplings beyond the edges;
comparison to periodic boundary conditions showed no discernible differences
for this lattice size at the temperatures examined below). and completely
random initial conditions, for $k_{\text{B}}T=$ 0.016, 0.032, 0.064,
0.128, 0.256 and 0.512 eV. We have used a non-square simulation lattice
because of the larger couplings in the $y$-direction compared to
the $x$-direction, which lead to longer correlation lengths in the
$y$-direction (see below). In \figref{SiZrO2_latt_MC_ss}, a sample
configuration of a well-thermalized simulation with $k_{\text{B}}T=0.016\ \text{eV}$
($T=186\ \text{K}$) is displayed. 

\begin{figure}
\begin{centering}
\includegraphics[width=0.8\columnwidth]{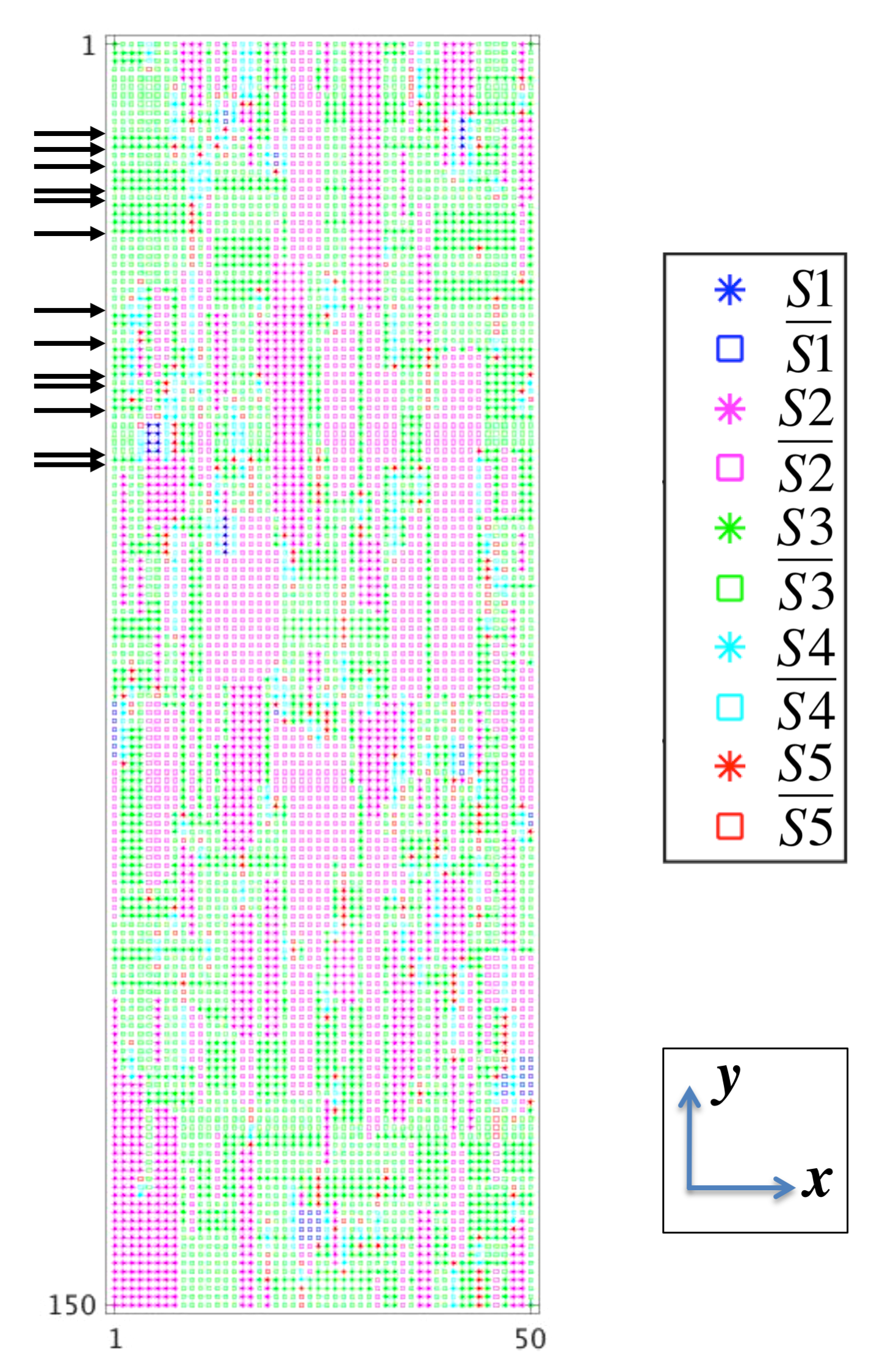}
\par\end{centering}
\caption{\label{fig:SiZrO2_latt_MC_ss} A snapshot of the Monte Carlo simulation
of the lattice model at $k_{\text{B}}T=0.016\ \text{eV}$ ($T=186\ \text{K}$).
On the left edge of the simulation frame, a series of domain walls
along the $x$-direction between $S3$ and $\overline{S3}$ domains
are emphasized by black arrows.}
\end{figure}

\begin{figure}
\begin{centering}
\includegraphics[width=1\columnwidth]{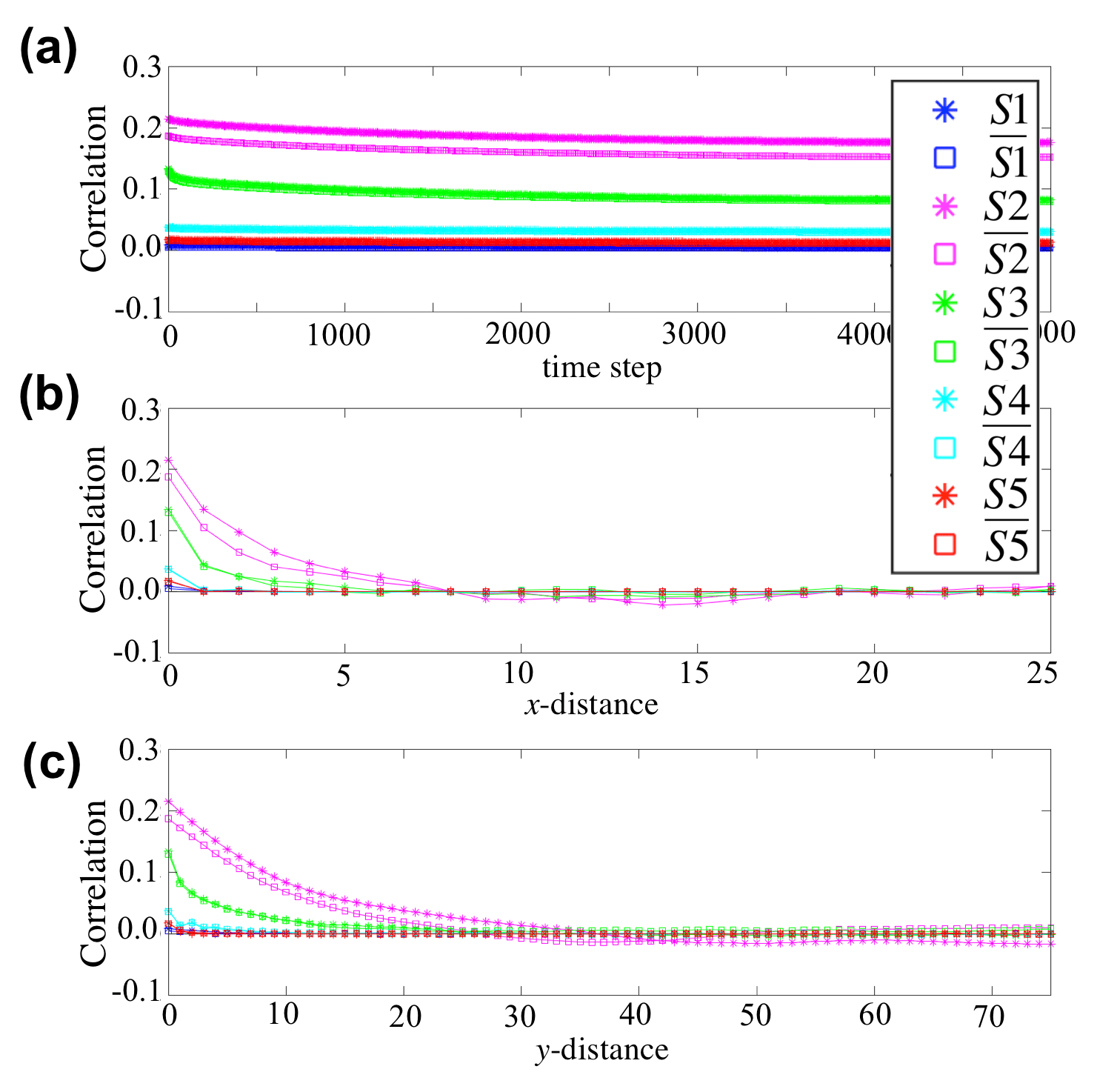}
\par\end{centering}
\caption[Temporal and spatial correlation functions of all 10 types of states
for a Monte Carlo simulation for $k_{\text{B}}T=0.016\ \text{eV}$.]{\label{fig:SiZrO2_latt_MC_corr}Temporal and spatial correlation
functions for all 10 states for a Monte Carlo simulation with $k_{\text{B}}T=0.016\ \text{eV}$.
(a) Temporal correlation (autocorrelation) functions as defined by
\eqref{CorTime}, (b) correlation functions along the $x$-direction
as per \eqref{CorX}, and (c) correlation functions along the $y$-direction
as per \eqref{CorY}. }
\end{figure}

In \figref{SiZrO2_latt_MC_corr}, the autocorrelation functions $C_{\text{auto}}^{(k)}\left(t\right)$
as a function of simulation step (``time'' $t$) and the horizontal
and vertical spatial correlation functions $C_{x}^{(k)}\left(\Delta i\right)$
and $C_{y}^{(k)}\left(\Delta j\right)$ are plotted for each state
$k$ for one particular Monte Carlo run. These correlation functions
are defined as
\begin{eqnarray}
C_{\text{auto}}^{\left(k\right)}\left(\Delta t\right) & = & \underset{i,j,t}{\text{mean}}\left[\left\langle \sigma_{k}\left(i,j,t\right)\sigma_{k}\left(i,j,t+\Delta t\right)\right\rangle \right.\nonumber \\
 &  & -\left.\left\langle \sigma_{k}\left(i,j,t\right)\right\rangle \left\langle \sigma_{k}\left(i,j,t+\Delta t\right)\right\rangle \right],\label{eq:CorTime}
\end{eqnarray}
\begin{eqnarray}
C_{x}^{\left(k\right)}\left(\Delta i\right) & = & \underset{i,j,t}{\text{mean}}\left[\left\langle \sigma_{k}\left(i,j,t\right)\sigma_{k}\left(i+\Delta i,j,t\right)\right\rangle \right.\nonumber \\
 &  & \left.-\left\langle \sigma_{k}\left(i,j,t\right)\right\rangle \left\langle \sigma_{k}\left(i+\Delta i,j,t\right)\right\rangle \right],\label{eq:CorX}
\end{eqnarray}
\begin{eqnarray}
C_{y}^{\left(k\right)}\left(\Delta j\right) & = & \underset{i,j,t}{\text{mean}}\left[\left\langle \sigma_{k}\left(i,j,t\right)\sigma_{k}\left(i,j+\Delta j,t\right)\right\rangle \right.\nonumber \\
 &  & \left.-\left\langle \sigma_{k}\left(i,j,t\right)\right\rangle \left\langle \sigma_{k}\left(i,j+\Delta j,t\right)\right\rangle \right],\label{eq:CorY}
\end{eqnarray}
where $\sigma_{k}\left(i,j,t\right)$ identifies the state at the
lattice site ($i,j$) at the simulation time step $t$. We have defined
10 functions $\sigma_{k}\left(i,j,t\right)$ (one for each state $k$)
such that $\sigma_{k}\left(i,j,t\right)=1$ if the lattice site ($i,j$)
is occupied by state $k$ at time $t$ and is 0 otherwise. In \figref{SiZrO2_latt_MC_ss},
correlation functions for every type of state ($S1$, $\overline{S1}$
etc.) are computed separately and overlaid.

We observe that for the run exemplified by \figref{SiZrO2_latt_MC_ss}
and analyzed in \figref{SiZrO2_latt_MC_ss}, (1) a 1000 step Monte
Carlo simulation leads to decorrelation (i.e., equilibration) of states
$S1$, $\overline{S1}$, $S4$, $\overline{S4}$, $S5$ and $\overline{S5}$
but not for $S2$, $\overline{S2}$, $S3$ and $\overline{S3}$. (2)
The simulation cell of size $50\times150$ is successful in containing
the domains that form at this temperature since the spatial correlations
become quite small by the half-way point along each direction of the
simulation cell: sites that are sufficiently far from each other are
not correlated. We have repeated these simulations 10 times for each
temperature and have found that the correlation functions behave similarly
when the initial state of the simulation cell is chosen randomly.
For temperatures higher than $0.128$ eV, all temporal correlations
decay below $0.1$ in the duration of the simulation. 

\begin{figure}
\begin{centering}
\includegraphics[width=0.9\columnwidth]{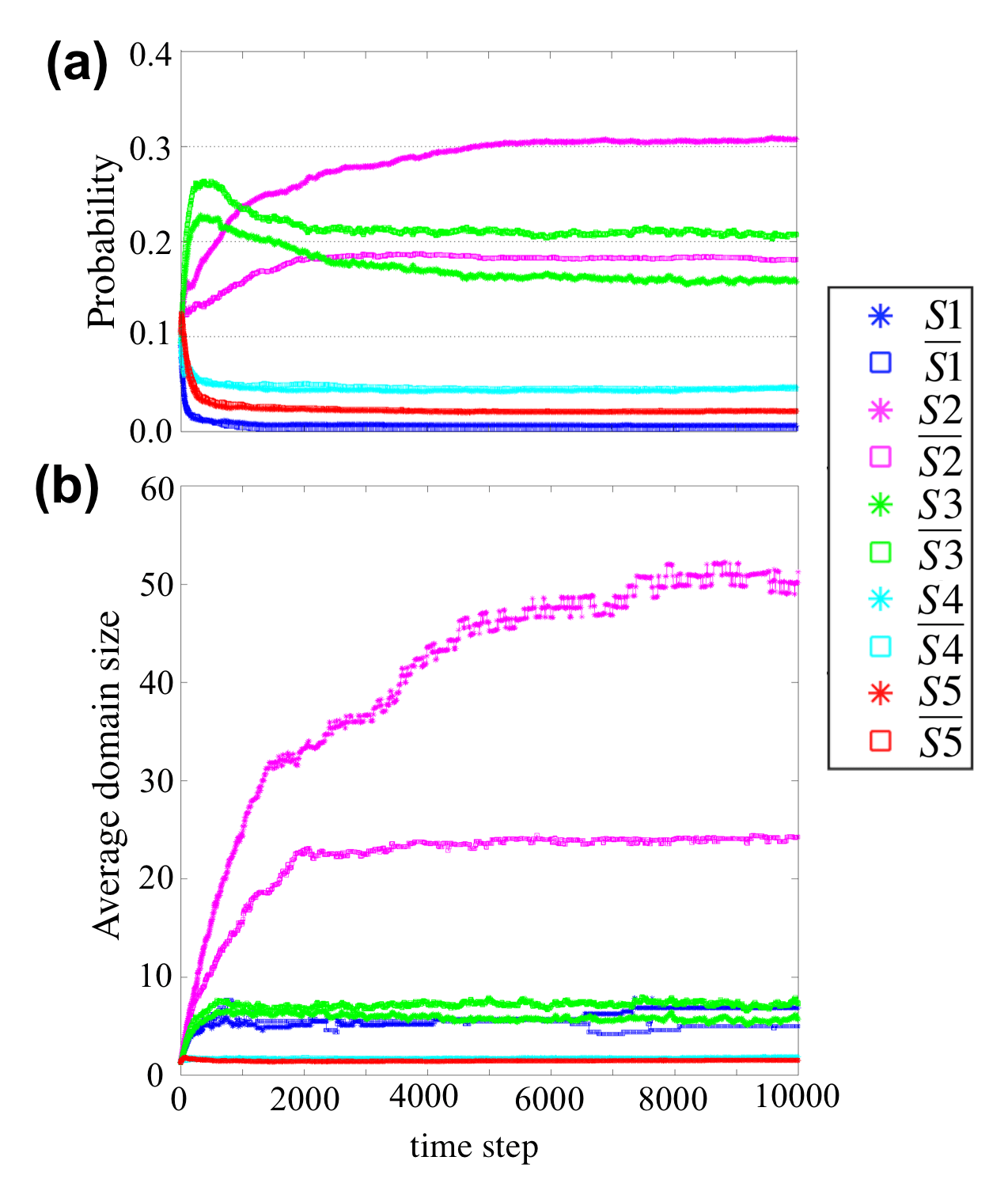}
\par\end{centering}
\caption[Probabilities of finding a state at an arbitrary site and average
domain sizes of each state, as they evolve during a Monte Carlo simulation
for $k_{\text{B}}T=0.016\ \text{eV}$.]{\label{fig:SiZrO2_latt_MC_other}Probabilities of finding a state
at an arbitrary site (a) and average domain sizes of each state (b),
as they evolve during a Monte Carlo simulation for $k_{\text{B}}T=0.016\ \text{eV}$.}
\end{figure}

The reason behind the slow temporal decay of the $S2$, $\overline{S2}$,
$S3$ and $\overline{S3}$ autocorrelations at low temperatures is
that large domains of these states form in the lattice, and the Monte
Carlo algorithm becomes inefficient in ``flipping'' these domains
to another configuration. To see what other effects are present in
these simulations, we monitor two other quantities displayed in \figref{SiZrO2_latt_MC_other}.
The first is the probability that any lattice site is occupied by
a particular state: we show the ratio of the number of sites occupied
by a particular state to the total number of sites in the simulation
cell. The second quantity is the average domain size for each state:
this is computed for each snapshot at a fixed time by first determining
all the domains of that state (including domains with only one site),
and then dividing the total number of sites occupied by the state
to the number of domains. A large jump in the second quantity during
the simulation usually indicates a merger of two domains. The fact
that these quantities change quickly at the beginning of the simulation
and more slowly toward the end of the simulation in \figref{SiZrO2_latt_MC_other}
is indicative that the characteristics seen in \figref{SiZrO2_latt_MC_ss}
are representative of large volumes of the configuration space sampled
with the Boltzmann distribution at $k_{\text{B}}T=0.016\ \text{eV}$
(186 K): namely, while the lattice system has not fully equilibrated,
i.e., the temporal correlations have not decayed to very small values,
it is not very far from equilibrium either. Hence, these results show
that at this low temperature, the lattice system should be dominated
by large domains of $S2$ and $\overline{S2}$ followed by smaller
domains of $S3$ and $\overline{S3}$.

\begin{figure}
\begin{centering}
\includegraphics[width=0.9\columnwidth]{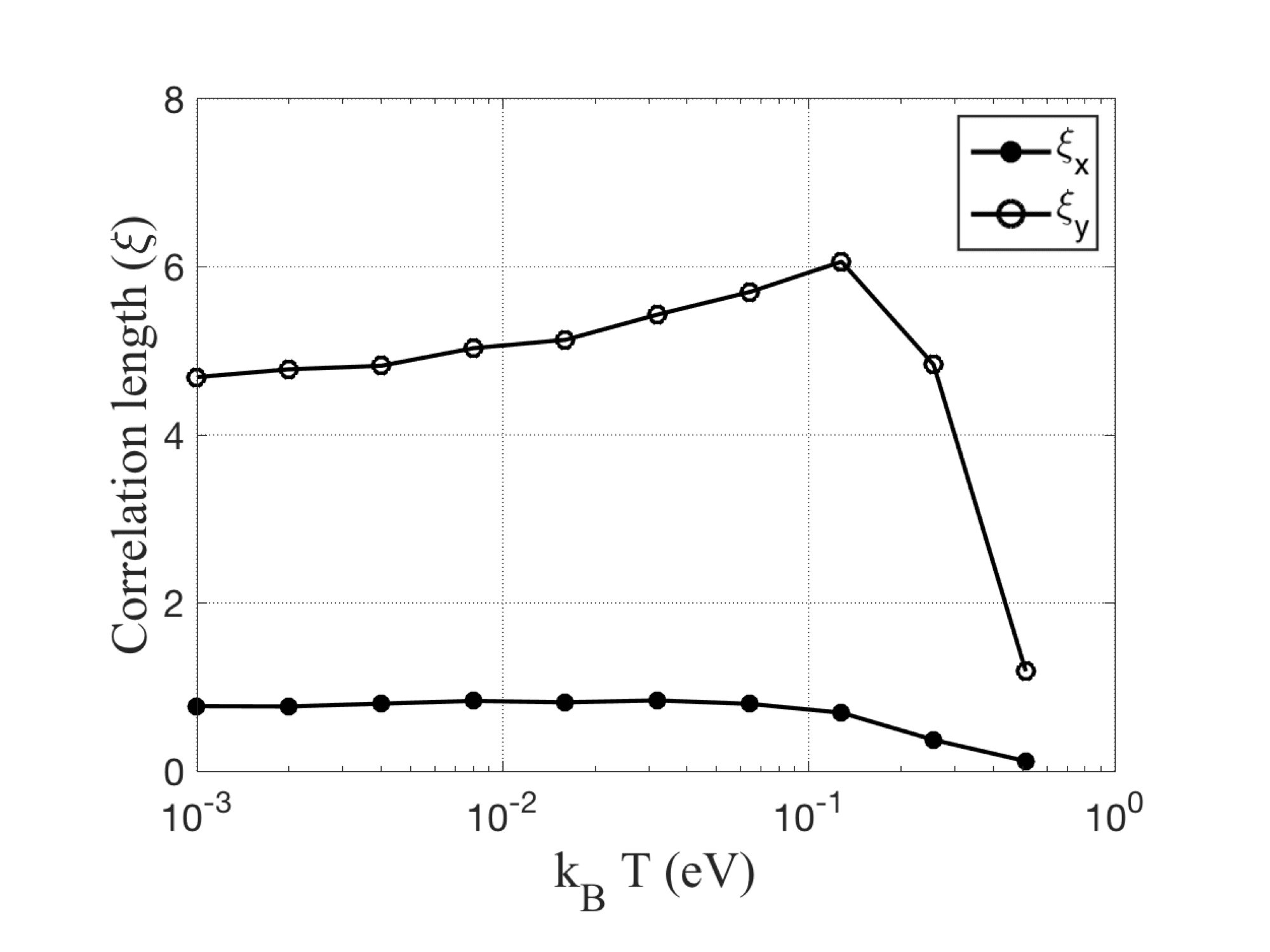}
\par\end{centering}
\caption[Correlation lengths along the $x$- and $y$-directions vs temperature.]{\label{fig:SiZrO2_latt_MC_CorLen}Correlation lengths along the $x$-
and $y$-directions vs temperature. Each data point is obtained by
fitting an exponential decay function to spatial correlation functions
for each run at a given temperature, and then averaging the results
of the fit for all the runs at that temperature.}
\end{figure}

We now return to the mean field prediction that at temperatures lower
than $1.4$ eV the system should be dominated by either one of the
ground states. Clearly, this prediction is not supported by our Monte
Carlo simulations. Our Monte Carlo simulations show that for $k_{B}T\gtrsim0.5$
eV, there is no long range order. In \ref{fig:SiZrO2_latt_MC_CorLen},
we plot the correlation lengths $\xi_{x}$ and $\xi_{y}$ along the
$x$- and $y$-directions, respectively. The correlation lengths are
calculated by fitting the spatial correlation functions $C_{x}^{\left(k\right)}\left(\Delta x\right)$
and $C_{y}^{\left(k\right)}\left(\Delta y\right)$ to exponentials
of the form $A\exp\left(-\Delta\alpha/\xi_{\alpha}\right)$. We calculate
the correlation lengths (averaged over all states) for each run and
then average over all runs at a given temperature. As indicated by
the temperature dependence of the correlation length $\xi_{y}$, the
system gradually becomes more ordered as the temperature is increased
up to $0.128\ \text{eV}$, and then becomes disordered. Such behavior
is associated with a second order phase transition in which correlation
lengths diverge upon approaching the critical temperature. If such
a critical temperature is present in this system, it lies between
$0.128\ \text{eV}$ ($\sim1500\ \text{K}$) and $0.256\ \text{eV}$
($\sim3000\ \text{K}$). Because the melting temperature of silicon
is $\sim1700\ \text{K}$, it is likely impossible to approach this
critical temperature in practice. Hence, it is safe to assume that
for the relevant experimental conditions ($T<1000\ \text{K}$), the
monolayer system is well within the ordered phase.

\begin{figure}
\begin{centering}
\includegraphics[width=1\columnwidth]{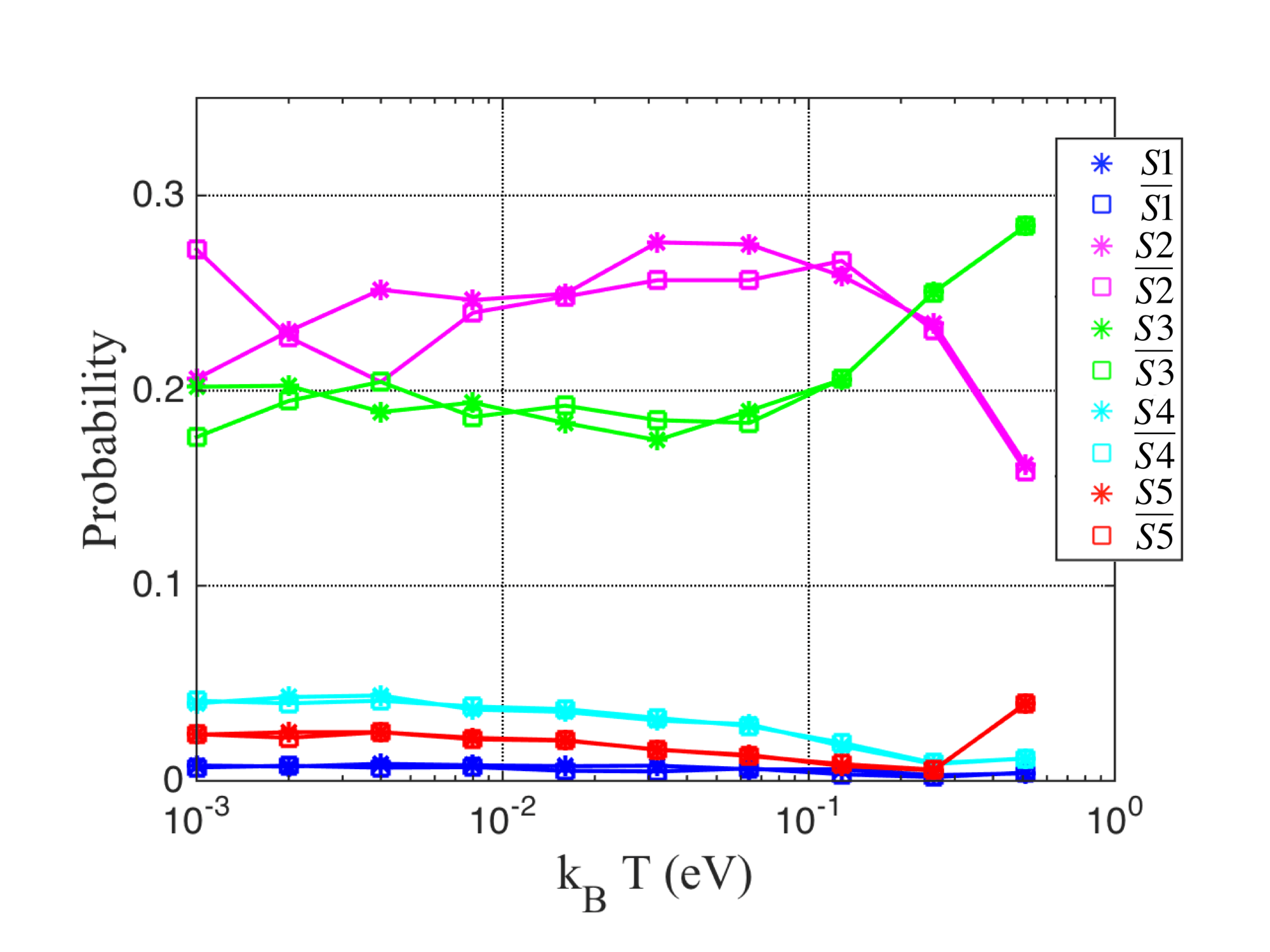}
\par\end{centering}
\caption[Probabilities of finding a state at an arbitrary site vs temperature,
as computed by Monte Carlo simulations.]{\label{fig:SiZrO2_latt_MC_Prob}Probabilities of finding a state
at an arbitrary site vs temperature, as computed by Monte Carlo simulations.
For each temperature, the probabilities are averaged over the last
quarter of each run, and then further averaged over 10 runs. }
\end{figure}

Finally, we comment on qualitative characteristics of the multi-domain
structure of these films based on our lattice model. In \figref{SiZrO2_latt_MC_Prob},
we display the probability for a site to be occupied by each state
as a function of temperature, where the probability values are averaged
over the last quarter of each run, and then further averaged over
10 runs. The data show that the system is dominated by the second
and the third lowest energy configurations ($S2,S3,\overline{S2},\overline{S3}$).
As discussed above, we believe that this is due to the rather low
couplings $J_{x}\left(S2,\overline{S3}\right)$ and $J_{y}\left(S3,\overline{S3}\right)$
when compared to the other couplings in \tabref{SiZrO2_latt_J}. Namely,
these domain walls are not very costly energetically, so their entropic
contribution is significant even at low temperatures and stabilizes
these phases even though they are not the lowest energy states.

\begin{figure}
\begin{centering}
\includegraphics[width=1\columnwidth]{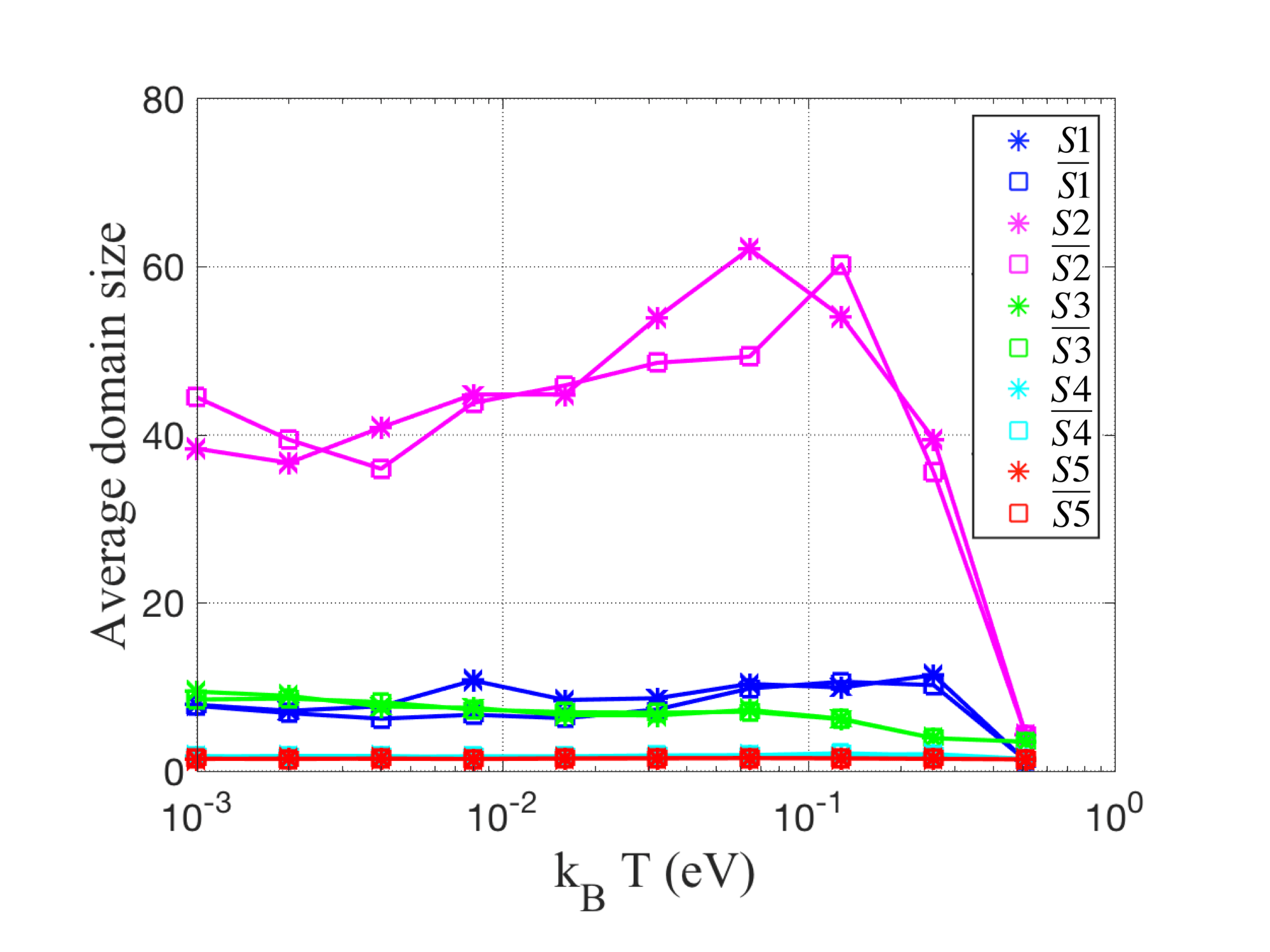}
\par\end{centering}
\caption[Average domain size for each type of state vs temperature, as computed
by Monte Carlo simulations.]{\label{fig:SiZrO2_latt_MC_Patches}Average domain size for each type
of state vs temperature, as computed by Monte Carlo simulations. For
each temperature, the domain sizes are averaged over the last quarter
of each run, and then further averaged over 10 runs.}
\end{figure}

In \figref{SiZrO2_latt_MC_Patches} we display the average domain
size of each state vs temperature, again averaged over 10 runs for
each temperature. We find that, on average, the domains of states
$S2$ and $\overline{S2}$ are larger than the domains of states $S3$
and $\overline{S3}$, even though they occupy similar portions of
the simulation cell (see \ref{fig:SiZrO2_latt_MC_Prob}). This may
be because $J_{y}\left(S3,\overline{S3}\right)=0.002\ \text{eV}$
so the $S3$ and $\overline{S3}$ easily form vertical stacks of domains
at essentially no energetic cost, as exemplified in \figref{SiZrO2_latt_MC_ss}:
some of these stacks are emphasized by black arrows on the left edge
of the figure, but there are many more in the interior of the simulation
cell.

To sum up, according to our discrete lattice model simulations, for
$2\times1$ ordered ZrO$_{2}$ monolayers on the Si(001) surface and
the experimentally relevant temperature range of $200-1000\ \text{K}$,
domains of $S2$, $\overline{S2}$, $S3$ and $\overline{S3}$ should
be expected to occur with linear extents ranging from a few to a few
dozen unit cells. This supports our claim that achieving epitaxy for
these films should be challenging. However, given that the local structure
is approximated by a mixture of $S2$ and $S3$ domains, the observed
ferroelectric switching is understandable as being due to a transition
between these two states.

\section{Conclusion\label{sec:Conclusion5}}

We have conducted a computational study of ZrO$_{2}$ monolayers o\textcolor{black}{n
Si(001) using DFT. These monolayers have recently been grown with
as an abrupt oxide/semiconductor interface but with an amorphous structure
and are measured to be ferroelectric \citep{dogan2018singleatomic}.
In our computations, we have found a multiplicity of (meta)stable
structures with a large variation in ionic polarization but small
differences in energy, atomic structure and chemistry. This suggests
that achieving epitaxy in the experiment should be challenging. In
order to understand the finite-temperature behavior of these ultrathin
films, we have developed a two dimensional discrete lattice model
of the domains in these thin films using DFT-derived parameters. We
have employed mean-field and Monte Carlo calculations to study this
lattice model and concluded that two distinct and oppositely polarized
structures, namely $S2$, $S3$ and their counterparts $\overline{S2}$
and $\overline{S3}$, dominate the films at the temperatures of interest.
The ferroelectric switching observed in the experiment is explained
by the film locally adopting one of these two structures and locally
switching between them. We have found that for monocrystalline epitaxial
films, this switching leads to a VBE shift in silicon of $\Delta V=0.6\ \text{eV}$,
which is moderately greater than the experimental value of $\Delta V=0.4\ \text{eV}$,
in agreement with the idea of partial (local) polarization switching.}

\section{Acknowledgements}

This work was supported primarily by the grant NSF MRSEC DMR-1119826.
We thank the Yale Center for Research Computing for guidance and use
of the research computing infrastructure, with special thanks to Stephen
Weston and Andrew Sherman. Additional computational support was provided
by NSF XSEDE resources via Grant TG-MCA08X007.

\bibliographystyle{apsrev}
\bibliography{Citations}

\end{document}


\chapter*{Supplementary Material for ``Theory of Ferroelectric ZrO$_{2}$
Monolayers on Si''}
\begin{center}
{\large{}Mehmet Dogan$^{1,2,3,4}$ and Sohrab Ismail-Beigi$^{1,2,5,6}$}\\
\par\end{center}

\begin{center}
$^{1}$Center for Research on Interface Structures and Phenomena,
Yale University, New Haven, Connecticut 06520, USA
\par\end{center}

\begin{center}
$^{2}$Department of Physics, Yale University, New Haven, Connecticut
06520, USA
\par\end{center}

\begin{center}
$^{3}$Department of Physics, University of California, Berkeley,
94720 USA
\par\end{center}

\begin{center}
$^{4}$Materials Science Division, Lawrance Berkeley National Laboratory,
Berkeley, California 94720, USA
\par\end{center}

\begin{center}
$^{5}$Department of Applied Physics, Yale University, New Haven,
Connecticut 06520, USA
\par\end{center}

\begin{center}
$^{6}$Department of Mechanical Engineering and Materials Science,
Yale University, New Haven, Connecticut 06520, USA
\par\end{center}

\section*{Monte Carlo algorithms for statistical physics}

A thermodynamic system is described by its partition function
\begin{equation}
Z=\sum_{\left\{ s\right\} }e^{-\frac{E_{s}}{k_{\text{B}}T}},
\end{equation}
where the sum runs over all possible states of the system, $E_{s}$
is the energy of state $s$, $k_{\text{B}}$ is Boltzmann's constant
and $T$ is the temperature. The expectation value of an observable
$X$ is
\begin{equation}
\left\langle X\right\rangle =\frac{1}{Z}\sum_{\left\{ s\right\} }X_{s}e^{-\frac{E_{s}}{k_{\text{B}}T}},
\end{equation}
where $X_{s}$ is the value of the observable $X$ when the system
is in state $s$.

The summations are over all possible states of the system which is
a space that is enormous for most physically relevant systems. However,
most of the states occur with vanishingly small probabilities, computed
by the formula $\frac{1}{Z}\exp\left(-\frac{E_{s}}{k_{\text{B}}T}\right)$.
Hence in order to avoid summing over all possible states, which is
an intractable problem and a wasteful attempt, one usually uses \emph{importance
sampling}, in which the sampling is done over states that are chosen
according to the probability distribution $\frac{1}{Z}\exp\left(-\frac{E_{s}}{k_{\text{B}}T}\right)$
\citep{luijten2006introduction}.

Given two states of the system and their energies, it is trivial to
compute their relative probabilities according to their Boltzmann
factors $\exp\left(-\frac{E_{s}}{k_{\text{B}}T}\right)$. However,
computing the absolute probability of a state requires computing $Z$,
which we wish to avoid. The most commonly used way of computing expectation
values without evaluating the partition function is by creating a
Markov chain of states in which each state only depends on the state
that immediately precedes it \citep{metropolis1953equation}. Starting
from a configuration $S_{i}$ with a Boltzmann factor $p_{i}$, a
new trial configuration $S_{j}$ with a Boltzmann factor $p_{j}$
is generated and accepted with probability $\pi_{ij}$. The probability
of occupying the state $S_{j}$ should be equal to the sum of the
probabilities of arriving at state $S_{j}$ from any given state $S_{i}$,
i.e. 
\begin{equation}
\sum_{i}p_{i}\pi_{ij}=p_{j}.
\end{equation}

At equilibrium, this Markov process should obey \emph{detailed balance},
i.e.
\begin{equation}
p_{i}\pi_{ij}=p_{j}\pi_{ji}.
\end{equation}

In general, the transition probabilities $\pi_{ij}$ are the product
of two factors: a probability $g_{ij}$ of proposing to move to state
$S_{j}$ from state $S_{i}$, and an acceptance ratio $A_{ij}$ of
accepting the proposed transition from $S_{i}$ to $S_{j}$. Thus
we can write
\begin{equation}
p_{i}g_{ij}A_{ij}=p_{j}g_{ji}A_{ji},
\end{equation}
or
\begin{equation}
\frac{g_{ij}A_{ij}}{g_{ji}A_{ji}}=\exp\left(-\frac{E_{j}-E_{i}}{k_{\text{B}}T}\right).\label{eq:ratios}
\end{equation}

For a given problem, $g_{ij},\ A_{ij}$ are specified by the algorithm
such that \Eqref{ratios} is satisfied and the sampling efficiency
is maximized.

Finally, a valid Monte Carlo algorithm must be ergodic, i.e., any
state must be reachable from any other state via a succession of moves.

\section*{Metropolis algorithms for discrete lattice models}

The most common Monte Carlo algorithm for discrete lattice models
such as the Ising model is the so-called Metropolis algorithm. Let
us describe this algorithm in the context of our lattice model which
we describe in more detail in the main text.

The Hamiltonian of our two dimensional discrete lattice model is
\begin{equation}
H=\sum_{i,j}E\left(\sigma\left(i,j\right)\right)+\sum_{i,j}J_{x}\left(\sigma\left(i,j\right),\sigma\left(i+1,j\right)\right)+\sum_{i,j}J_{y}\left(\sigma\left(i,j\right),\sigma\left(i,j+1\right)\right),
\end{equation}
where $\left(i,j\right)$ are the positions on the discrete lattice
along the $\left(x,y\right)$-directions, $\sigma\left(i,j\right)$
is the state on lattice site $\left(i,j\right)$, $E\left(\sigma\left(i,j\right)\right)$
is the site energy of the state $\sigma\left(i,j\right)$, $J_{x}\left(\sigma\left(i,j\right),\sigma\left(i+1,j\right)\right)$
is the nearest-neighbor interaction energy along the $x$-direction,
and $J_{y}\left(\sigma\left(i,j\right),\sigma\left(i,j+1\right)\right)$
is the nearest-neighbor interaction energy along the $y$-direction.
$J_{x}\left(\sigma_{1},\sigma_{2}\right)=J_{y}\left(\sigma_{1},\sigma_{2}\right)=0$
if $\sigma_{1}=\sigma_{2}$. In this model, there are $N$ types of
states, i.e. $\sigma$ is a function that maps a lattice site onto
one of $s_{1},s_{2},\ldots,s_{N}$. Note that the lower-case $s$
are different from the upper-case $S$ used above, which denoted the
state of the whole system, which would be the collection of states
$s$ on all lattice points for this model.

The two dimensional Ising model is a special case of our model, where
$N=2$. The external magnetic field can be included by having $E\left(s_{1}\right)\neq E\left(s_{2}\right)$,
and anisotropy can be included by having $J_{x}\left(s_{1},s_{2}\right)\neq J_{y}\left(s_{1},s_{2}\right)$.

The Metropolis algorithm would operate on our $N$-state model as
follows:
\begin{enumerate}
\item Pick a lattice site at random. Let us call the state on the site $s_{i}$.
Let us call the state of the initial system $S_{\mu}$.
\item Propose to flip the state $s_{i}$ to another state $s_{f}$, chosen
among all non-$s_{i}$ states with equal probability $\frac{1}{N-1}$.
Let us call the state of the system if the proposed flip occurs $S_{\nu}$.
Thus $g_{\mu\nu}=\frac{1}{N-1}$ (see \Eqref{ratios}). The probability
of proposing the inverse move, i.e. going to $S_{\mu}$ from $S_{\nu}$
is clearly the same, hence $g_{\nu\mu}=g_{\mu\nu}=\frac{1}{N-1}$.
\item Compute the energy difference $E_{\nu}-E_{\mu}$ between $S_{\nu}$
and $S_{\mu}$. This is simple, since the only difference is the state
change of state $s_{i}$ to $s_{j}$, and the energy difference is
localized to the site energy and the couplings with the nearest neighbors
of that site.
\item The acceptance ratios are obtained by \Eqref{ratios}:
\begin{equation}
\frac{A_{\mu\nu}}{A_{\nu\mu}}=\exp\left(-\frac{E_{\nu}-E_{\mu}}{k_{\text{B}}T}\right).
\end{equation}
A common way of achieving this equation is by setting:
\begin{equation}
A_{\mu\nu}=\begin{cases}
\exp\left(-\frac{E_{\nu}-E_{\mu}}{k_{\text{B}}T}\right) & \text{if}\ E_{\nu}>E_{\mu}\\
1 & \text{if}\ E_{\nu}\leq E_{\mu}
\end{cases}\label{eq:Accept1}
\end{equation}
\end{enumerate}
To find the expectation value of an observable $X$, $X$ is computed
at each step of the simulation that comprises of a finite number of
steps, and then simply averaged. This is the merit of \emph{importance
sampling}, which takes care of the relative probabilities of states
through \Eqref{ratios}, therefore the observables can simply be averaged.

\section*{Wolff cluster algorithms}

The success of a Monte Carlo algorithm is usually measured by how
easy it can generate ``independent'' samples, i.e. how many attempts
it takes to go from a state $S_{\mu}$ to another state $S_{\nu}$
such that $S_{\mu}$ and $S_{\nu}$ are ``uncorrelated'' (namely,
decorrelation time). The ``single-flip'' Metropolis algorithm is
conceptually simple and easy to implement. However, at each simulation
step, the state only slightly changes, so the decorrelation time can
be large. For models with a second order phase transition, such as
the two dimensional Ising model, this algorithm suffers from ``critical
slowing down'' where, close to the critical temperature of the model,
the decorrelation time diverges \citep{barkema1997newmonte}.

This issue can be solved by algorithms that propose states that are
sufficiently modified from the preceding state. A family of such algorithms
is called cluster algorithms, where rather than switching the state
on a single site, the state on a groups of sites (``a cluster'')
is switched simultaneously \citep{swendsen1987nonuniversal}. Here
we modify the Wolff cluster algorithm \citep{wolff1989collective},
originally developed for the Ising model, to simulate our $N$-state
model:

\begin{figure}
\begin{centering}
\includegraphics[width=0.75\columnwidth]{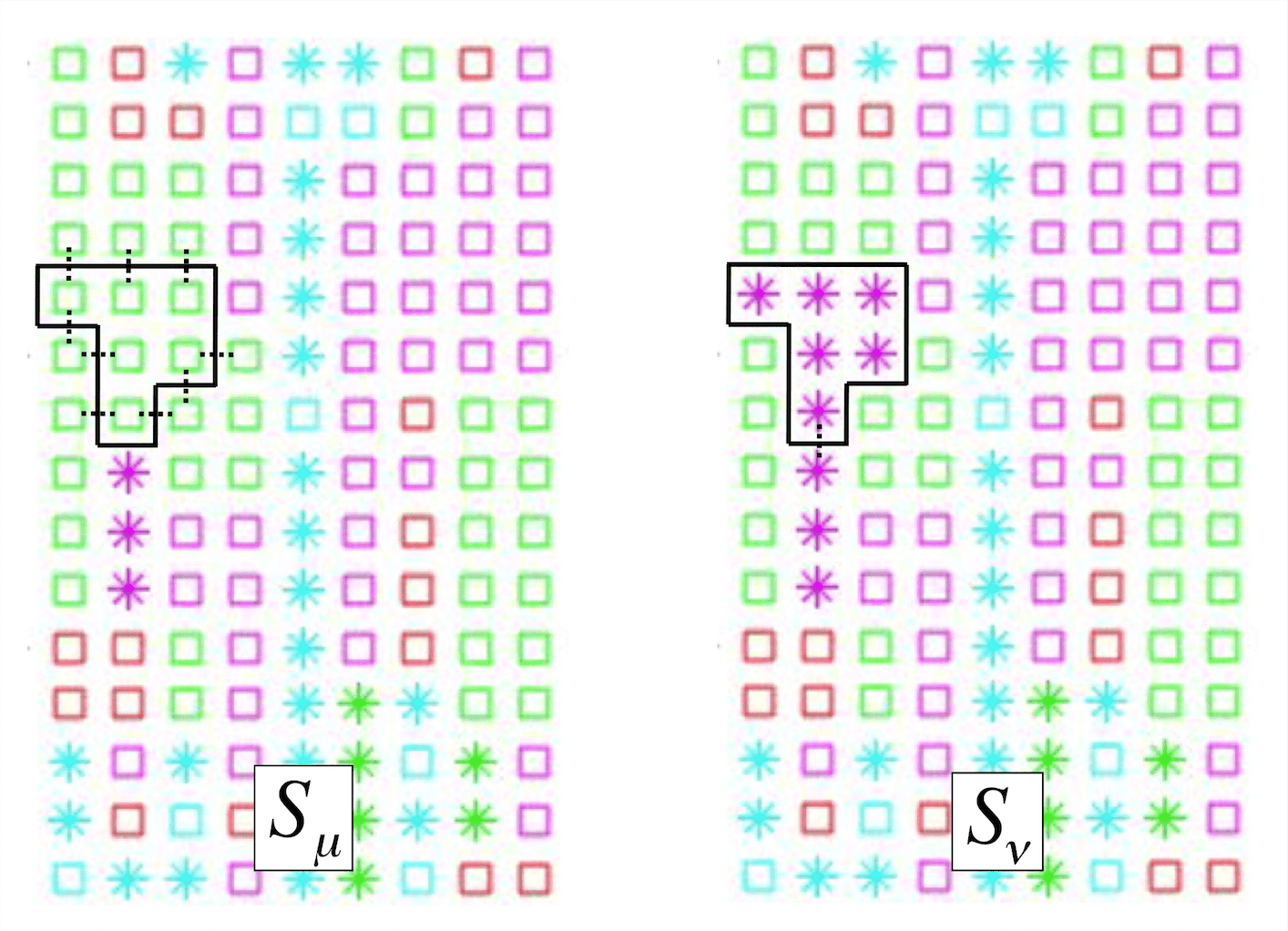}
\par\end{centering}
\caption[A sample instant of a Wolff cluster simulation of an -state lattice
model, prior to and after the switching of a cluster.]{\label{fig:Wolff}A sample instant of a Wolff cluster simulation
of an $N$-state lattice model, prior to ($S_{\mu}$) and after ($S_{\nu}$)
the switching of a cluster. The boundary of the cluster is shown by
solid lines, and the bonds at the boundary of the cluster are shown
by dotted lines. Each color-shape combination denotes a type of state
in our 10-state lattice model, described in detail in the main text.}
\end{figure}

\begin{enumerate}
\item Pick a lattice site $i$ at random. Let us call the state on the site
$s_{i}$. Let us call the state of the initial system $S_{\mu}$.
\item Add each of the nearest neighbors $j$ of the site $i$ to the cluster,
with the probability $p_{\text{add}}$, provided that the states on
sites $i$ and $j$ are the same, and the ``bond'' between $i$
and $j$ has not yet been considered.
\item Once all the neighbors of site $i$ have been considered, move to
the next site in the cluster. Repeat step 2 for this site. If all
the sites in the cluster have gone through step 2, the cluster has
been built. Move to step 4.
\item Propose to flip the state $s_{i}$ to another state $s_{f}$, chosen
among all non-$s_{i}$ states with equal probability $\frac{1}{N-1}$,
for all the sites in the cluster. Let us call the state of the system
if the proposed flip occurs $S_{\nu}$.
\item Compute the number of bonds at the boundary of the cluster. The two
neighboring states of the same kind are said to have a bond that is
intact. When the cluster is ``flipped'' the bonds at the boundary
will be broken. In \Figref{Wolff}, we illustrate the formation of
a cluster for a given state $S_{\mu}$ of the lattice, shown on the
left. The number of bonds at the boundary (shown as dotted lines in
the figure) is $n_{\mu}=9$. The proposed state $S_{\nu}$ is shown
on the right. The number of bonds at the boundary in the proposed
state is $n_{\nu}=1$.
\item Compute the energy difference $E_{\nu}-E_{\mu}$ between $S_{\nu}$
and $S_{\mu}$. This requires accounting for all the nearest-neighbor
interactions at the boundary of the cluster in both the initial and
the final states.
\end{enumerate}
Finding the correct acceptance ratio for this algorithm is somewhat
involved. Let us assume that the cluster in $S_{\mu}$ in \Figref{Wolff}
is built in the following order:
\begin{enumerate}
\item The site at the upper left corner of the cluster is randomly picked.
\item The site to the right is added with probability $p_{\text{add}}$,
the other neighboring sites of the same kind (above and below) are
rejected with probability $\left(1-p_{\text{add}}\right)^{2}$.
\item The site to the right is added with probability $p_{\text{add}}$,
the other neighboring sites of the same kind (above and below) are
rejected with probability $\left(1-p_{\text{add}}\right)^{2}$.
\item The site below is added with probability $p_{\text{add}}$, the site
above is rejected with probability $\left(1-p_{\text{add}}\right)$.
\item The site to the left is added with probability $p_{\text{add}}$,
the other neighboring sites of the same kind (to the right and below)
are rejected with probability $\left(1-p_{\text{add}}\right)^{2}$.
\item The site below is added with probability $p_{\text{add}}$, the site
to the left is rejected with probability $\left(1-p_{\text{add}}\right)$.
\item Both neighboring sites (to the left and to the right) are rejected
with probability $\left(1-p_{\text{add}}\right)^{2}$.
\end{enumerate}
The total probability of this process in this order is $p_{\text{add}}^{5}\left(1-p_{\text{add}}\right)^{10}$.
The same process can be repeated for the cluster in $S_{\nu}$ in
\Figref{Wolff} built in the exact same order, which yields a probability
of $p_{\text{add}}^{5}\left(1-p_{\text{add}}\right)^{2}$.

The ratio of proposal probabilities of the forward and backward moves
is then
\begin{equation}
\frac{p_{\text{add}}^{5}\left(1-p_{\text{add}}\right)^{10}}{p_{\text{add}}^{5}\left(1-p_{\text{add}}\right)^{2}}=\left(1-p_{\text{add}}\right)^{8},
\end{equation}
where 8 is the difference in the number of bonds at the boundary for
$S_{\mu}$ and $S_{\nu}$, i.e. $n_{\mu}-n_{\nu}=8$. It is evident
that for any given order for building the same cluster, the ratio
of proposal probabilities of the forward and backward moves will be
$\left(1-p_{\text{add}}\right)^{n_{\mu}-n_{\nu}}$. Because $g_{\mu\nu}$
is the sum of the probabilities of all moves that propose $S_{\nu}$
from $S_{\mu}$ and $g_{\nu\mu}$ is the sum of the probabilities
of all moves that propose $S_{\mu}$ from $S_{\nu}$, we can write
\begin{equation}
\frac{g_{\mu\nu}}{g_{\nu\mu}}=\left(1-p_{\text{add}}\right)^{n_{\mu}-n_{\nu}}.
\end{equation}

Therefore \Eqref{ratios} yields
\begin{align}
\frac{A_{\mu\nu}}{A_{\nu\mu}} & =\left(1-p_{\text{add}}\right)^{n_{\nu}-n_{\mu}}\exp\left(-\frac{E_{\nu}-E_{\mu}}{k_{\text{B}}T}\right)\nonumber \\
 & =\exp\left(-\frac{E_{\nu}-E_{\mu}-k_{\text{B}}T\left(n_{\nu}-n_{\mu}\right)\log\left(1-p_{\text{add}}\right)}{k_{\text{B}}T}\right).
\end{align}

If we define
\begin{equation}
\Delta_{\mu\nu}\equiv E_{\nu}-E_{\mu}-k_{\text{B}}T\left(n_{\nu}-n_{\mu}\right)\log\left(1-p_{\text{add}}\right),\label{eq:Delta}
\end{equation}
we can set the acceptance ratios (in analogy with \Eqref{Accept1})
to be
\begin{equation}
A_{\mu\nu}=\begin{cases}
\exp\left(-\frac{\Delta_{\mu\nu}}{k_{\text{B}}T}\right) & \text{if}\ \Delta_{\mu\nu}>0\\
1 & \text{if}\ \Delta_{\mu\nu}\leq0
\end{cases}.\label{Accept2}
\end{equation}

In the original Wolff cluster method for the Ising model, $p_{\text{add}}$
is defined as a function of temperature such that the acceptance ratios
are always 1. This makes for a rejection-less algorithm which is able
to switch clusters of different sizes at any temperature. However,
in our model there is no simple relationship between $E_{\nu}-E_{\mu}$
and $n_{\nu}-n_{\mu}$ as in there is in the Ising model. Therefore
$p_{\text{add}}$ cannot be defined \emph{a priori} to make $\Delta_{\mu\nu}$
vanish in \Eqref{Delta}, which in turn would guarantee $A_{\mu\nu}=1$
in \ref{Accept2}. After empirical tests on our simulations, we have
set $p_{\text{add}}=\frac{1}{2}$ for the results presented in the
main text. Improving the acceptance ratios through the choice of $p_{\text{add}}$
is the subject of future research.

\section*{List of all domain wall energies}

We tabulate all domain wall energies in \tabref{SiZrO2_all_J}, which
includes the couplings between $S1$, $S2$, $S3$ and their barred
counterparts, also reported above in \tabref{SiZrO2_latt_J}.

\begin{table}
\begin{centering}
\begin{tabular}{ccc}
\toprule 
\addlinespace[0.3cm]
Domain boundary & $\ \ J_{x}$ (eV)$\ \ $ & $\ \ J_{y}$ (eV)$\ \ $\tabularnewline\addlinespace[0.3cm]
\midrule
\addlinespace[0.1cm]
\midrule 
$S1,\overline{S1}$ & 0.26 & 1.35\tabularnewline
\midrule 
$S1,S2$ & 0.76 & 1.13\tabularnewline
\midrule 
$S1,\overline{S2}$ & 0.96 & 0.99\tabularnewline
\midrule 
$S1,S3$ & 0.61 & 4.81\tabularnewline
\midrule 
$S1,\overline{S3}$ & 0.44 & 1.75\tabularnewline
\midrule 
$S1,S4$ & 0.55 & 2.79\tabularnewline
\midrule 
$S1,\overline{S4}$ & -0.20 & 2.37\tabularnewline
\midrule 
$S1,S5$ & 0.35 & 1.35\tabularnewline
\midrule 
$S1,\overline{S5}$ & 0.40 & 0.56\tabularnewline
\midrule 
$S2,\overline{S2}$ & 0.38 & 1.64\tabularnewline
\midrule 
$S2,S3$ & 0.17 & 0.98\tabularnewline
\midrule 
$S2,\overline{S3}$ & 0.01 & 0.91\tabularnewline
\midrule 
$S2,S4$ & -0.17 & 2.23\tabularnewline
\midrule 
$S2,\overline{S4}$ & 0.86 & 2.28\tabularnewline
\midrule 
$S2,S5$ & 0.34 & 0.59\tabularnewline
\midrule 
$S2,\overline{S5}$ & -0.12 & 1.31\tabularnewline
\midrule 
$S3,\overline{S3}$ & 0.73 & 0.002\tabularnewline
\midrule 
$S3,S4$ & 0.10 & 1.89\tabularnewline
\midrule 
$S3,\overline{S4}$ & 0.23 & 1.84\tabularnewline
\midrule 
$S3,S5$ & -0.24 & 0.75\tabularnewline
\midrule 
$S3,\overline{S5}$ & 0.69 & 1.21\tabularnewline
\midrule 
$S4,\overline{S4}$ & 0.55 & -0.33\tabularnewline
\midrule 
$S4,S5$ & 0.71 & 0.65\tabularnewline
\midrule 
$S4,\overline{S5}$ & -0.26 & 1.86\tabularnewline
\midrule 
$S5,\overline{S5}$ & 0.56 & 0.30\tabularnewline
\bottomrule
\end{tabular}
\par\end{centering}
\caption[Domain boundary energies computed from first principles.]{\label{tab:SiZrO2_all_J}Domain boundary energies computed from first
principles. These energies serve as the couplings of nearest neighbors
in our lattice model.}
\end{table}

\begin{figure}
\begin{centering}
\includegraphics[width=0.9\columnwidth]{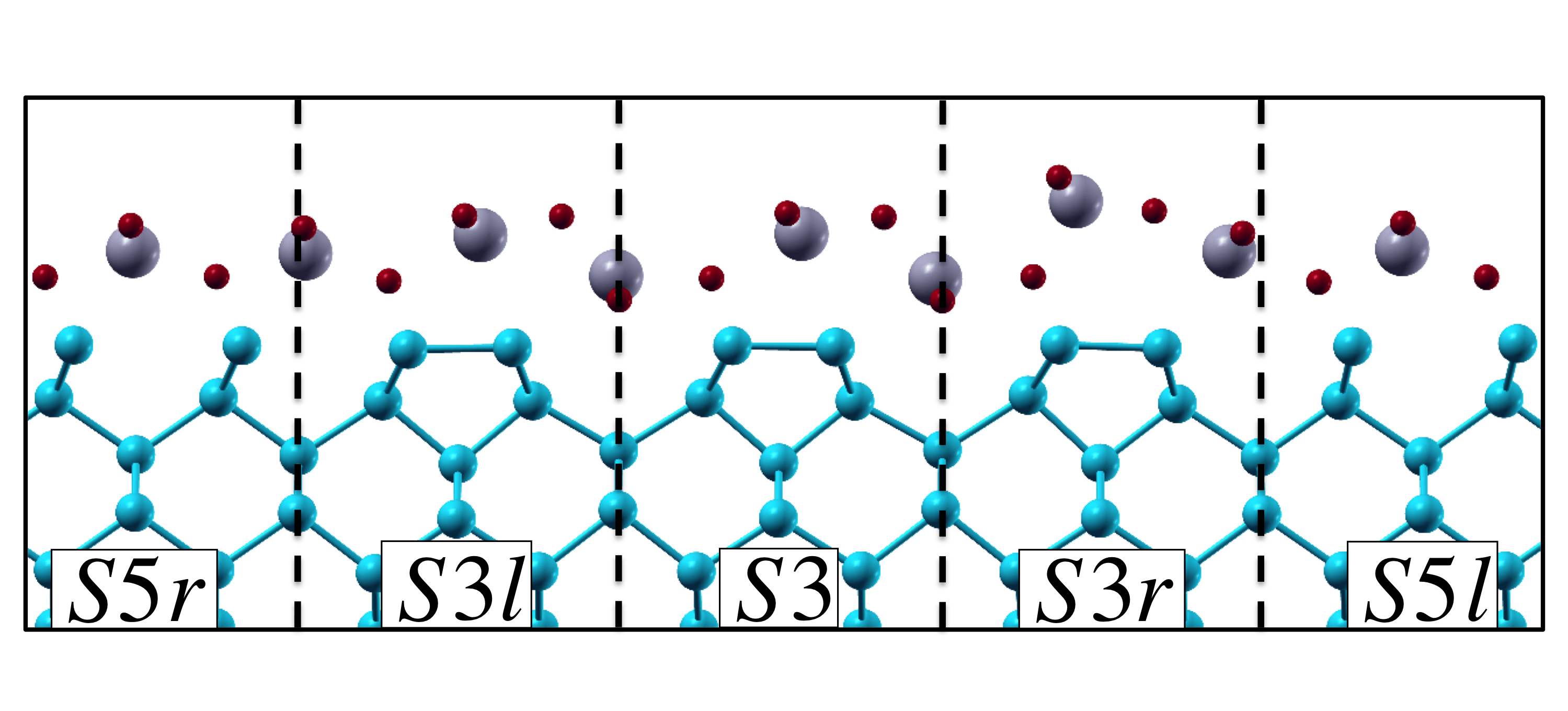}
\par\end{centering}
\caption[The domain boundaries along the $y$-direction between $S3$ and $S5$,
computed by stacking 3 unit cells of each structure in the $x$-direction.]{\label{fig:SiZrO2_dom_S3S5}The domain boundaries along the $y$-direction
between $S3$ and $S5$, computed by stacking 3 unit cells of each
structure along the $x$-direction. The domain energy, computed to
be $J_{x}\left(S5,S5\right)=-0.24\ \text{eV}$ per unit length, is
negative partly due to the fact that the vicinity of one of the boundary
walls (labelled $S3r$) resembles a lower energy configuration $S2$
(see Figure 5 in the main text).}
\end{figure}

A few of the couplings that involve the higher-energy $S4$ and $S5$
structures are negative, which can be understood in some cases when
the domain boundary region resembles a lower energy structure. In
\figref{SiZrO2_dom_S3S5}, we illustrate the domain boundaries that
correspond to $J_{x}\left(S3,S5\right)=-0.24\ \text{eV}$. The structure
immediately to the left of the right domain boundary ($S3r$) closely
resembles the $S2$ structure (see Figure 5 in the main text). However,
the fact that the higher energy $S4$ and $S5$ structures have negative
domain wall energies with the lower energy structures in some cases
is not enough to generate antiferroelectric patterns in our Monte
Carlo simulations. This may be due to the separation of scale in the
energies of the lowest three structures and $S4$ and $S5$ (see Table
II in the main text). Hence the energy reduction achieved by making
these domain boundaries (0.26 eV or less) is not enough to compensate
for the high energy cost of creating these two structures in the first
place.

\bibliographystyle{apsrev}
\bibliography{Citations-sup}